\documentclass[%
 aip,
 amsmath,amssymb,
 reprint,%
]{revtex4-1}

\usepackage{graphicx}
\usepackage{dcolumn}
\usepackage{bm}

\usepackage[utf8]{inputenc}
\usepackage[T1]{fontenc}
\usepackage{mathptmx}
\usepackage{enumerate}
\usepackage{bbm}
\usepackage{xcolor}
\usepackage{float}
\usepackage{amsfonts}
\usepackage{amssymb}
\usepackage{dsfont}
\usepackage{enumerate}

\graphicspath{{./../Figures/}}

\newcommand{\ve}{\vec{e}}
\newcommand{\vF}{\vec{F}}

\newcommand{\vG}{\vec{G}}
\newcommand{\vR}{\vec{R}}
\newcommand{\vV}{\vec{V}}
\newcommand{\vX}{\vec{X}}
\newcommand{\vVa}{\vec{V}^\ast}
\newcommand{\vS}{\vec{S}}
\newcommand{\vu}{\vec{u}}
\newcommand{\vxi}{\vec{\xi}}
\newcommand{\vdel}{\vec{\delta} V}
\newcommand{\veta}{\vec{\eta}}

\newcommand\cA{{\mathcal A}}
\newcommand\cD{{\mathcal D}}
\newcommand\cI{{\mathcal I}}
\newcommand\cF{{\mathcal F}}
\newcommand\cG{{\mathcal G}}
\newcommand\cJ{{\mathcal J}}
\newcommand\cM{{\mathcal M}}
\newcommand\cN{{\mathcal N}}

\newcommand\phisp[1]{\phi^{(eq)}\pare{#1}}\newcommand{\pare}[1]{\left(\, #1 \, \right)}
\newcommand{\brk}[1]{\langle\, #1  \, \rangle}

\newcommand{\bra}[1]{\left[\, #1 \, \right]}

\newcommand{\Set}[1]{\left\{\, #1 \, \right\}}

\newcommand{\Prob}[1]{P\left[\, #1 \, \right]}
\newcommand{\Probc}[2]{P\bra{#1 \, \left| \, #2 \right.}}
\newcommand\seq[3]{{#1}_{#2}^{#3}}
\newcommand\gk[1]{g_k\pare{#1}}
\newcommand{\deq}{\stackrel {\rm def}{=}}
\newcommand\tko{\tau_k(t,\omega)}

\newcommand\Vkspont[1]{V_k^{(eq)}({#1})}
\newcommand\VkL[1]{V_k^{(L)}({#1})}

\newcommand\Vksyn[1]{V_k^{(syn)}({#1})}

\newcommand\Vkext[1]{V_k^{(S)}({#1})}
\newcommand\Vkdet[1]{V_k^{(det)}({#1})}
\newcommand\Vknoise[1]{V_k^{(noise)}({#1})}

\newcommand{\Probct}[3]{P_{#1}\bra{#2 \, \left| \, #3 \right.}}
\newcommand\bloc[2]{{\omega}_{#1}^{#2}}
\newcommand\sk[1]{\sigma_k({#1})}

\newcommand\moy[1]{\mu\bra{#1}}
\newcommand{\setN}{\mathbbm{N}}

\newcommand{\setR}{\mathbbm{R}}
\newcommand{\ent}[1]{\left[\, #1 \, \right]}

\newcommand\nkrm{\tau_k(r-1,\omega)}
\newcommand\dmu[1]{\delta^{(1)}\moy{#1}}
\newcommand\tLk{\tau_{L,k}}
\newcommand\skd[1]{\sigma^2_k({#1})}
\newcommand\Csp[2]{\mathcal{C}^{(eq)}\bra{#1, #2}}
\newcommand{\vect}[1]{\pare{\begin{array}{ccc}#1\end{array}}}

\newcommand{\abs}[1]{\left|\, #1 \, \right|}
\newcommand{\Pnc}[2]{\mathds{P}_n\bra{#1 \, \left| \, #2 \right.}}
\newcommand\sif[1]{\seq{\omega}{-\infty}{#1}}
\newcommand{\connect}[3]{#1 \, \stackrel{#3}{\rightarrow} \, #2}
\newcommand\musp{\mu^{(eq)}}
\newcommand\moysp[1]{\mu^{(eq)}\bra{#1}}
\newcommand{\Expm}[2]{\mathds{E}_{#1}\left[\, #2 \, \right]}

\newcommand{\beq}{\begin{equation}}
\newcommand{\eeq}{\end{equation}}

\begin{document}


\title[Linear response in neuronal 
networks]{Linear response in neuronal 
networks: \\
from neurons 
dynamics to collective 
response}

\author{B. Cessac}
 \altaffiliation[]{Université Côte d'Azur, Inria, Biovision team, France
}%

\date{\today}

\begin{abstract}
We review two examples where the linear response of a neuronal network submitted to an external stimulus can be derived explicitly, including network parameters dependence. This is done in a statistical physics-like approach where one associates to the spontaneous dynamics of the model a natural notion of Gibbs distribution inherited from ergodic theory or stochastic processes. These two examples are the Amari-Wilson-Cowan model \cite{amari:72,wilson-cowan:72} and a conductance based Integrate and Fire model  \cite{rudolph-destexhe:06,rudolph-destexhe:07}.  
\end{abstract}

\maketitle

\begin{quotation}
One of the contemporary challenges in  neuroscience is to understand how our brain processes external world information. For example, our retina receives the light coming from a visual scene and efficiently converts it into trains of impulses (action potentials) sent to the brain via the optic nerve. The visual cortex is then able to decode this flow of information in a fast and efficient way. How does a neuronal network, like the retina, adapts its internal dynamics to stimuli, yet providing a response that can be successfully deciphered by another neuronal network ? Even if this question is far from being resolved, there exist successful methods and strategies providing partial answers. To some extent, as developed in this paper, this question can be addressed from the point of non equilibrium statistical physics and  linear response theory. Although neuronal networks are  outside the classical scope of non equilibrium statistical physics - interactions (synapses) are not symmetric, equilibrium evolution is not time-reversible, there is no known conserved quantity, no Lyapunov function -  an extended notion of Gibbs distribution can be proposed, directly constructed from the dynamics where the linear response can be derived explicitly, including network parameters dependence.
\end{quotation}

\section{Introduction}\label{sec:Introduction}

Our nervous system has the capacity to display adapted responses to its environment, in a fast way, e.g. compatible with survival, with low energy consumption so as to maintain the body temperature in a narrow range. 
The main cells involved in this process are the neurons, although other non neuronal cells, like glia, play a central role too,
(see the website \url{http://www.networkglia.eu/en/classicpapers}). Our brain is made of neuronal networks with specific functions.
A central question in neuroscience is to understand how 
these neuronal networks contribute to encode and decode the external world information so as to be able to respond in a fast and adapted way. Behind this question is the hope to be able, one day, to decipher the "neural code" \cite{rieke-warland-etal:97}, so as, for example, to stimulate the impaired retina of a patient in a proper way (e.g. with a retinal prosthesis) to partially restore his vision (see e.g. \cite{tran:15} and references therein). 

Although we are yet far from this, it is already possible to characterise the response of sensory neurons to stimuli (e.g. in the retina or in the visual cortex) . The first step of this characterisation uses a linear response approach, where the response $R$ to a stimuli is approximated by a spatio-temporal response kernel $K$, convolved with the stimulus $S$, $R \sim K \ast S$. This kernel can be empirically computed from the cell's activity, thereby allowing the deconvolution of the signal to find back the stimulus \cite{chichilnisky:01,pillow-paninski-etal:05}.
 In the field of vision, the spatio-temporal kernel $K$ characterises the receptive field of a sensory neuron and is a fundamental building element in Hubel and Wiesel's theory of visual perception \cite{hubel-wiesel:62}.   
It is known that the receptive field $K$ does not only results from the intrinsic properties of the cell, but from its connections with other neurons. It involves therefore network effects. It is also known that the linear response approach is in general not sufficient to fully characterise the neurons' response property. It is only a first, essential step, but non linear corrections are usually necessary.
 
From a more fundamental point of view, it is not established yet how the non linear neurons' and synapses' dynamics, the synaptic architecture, the influence of noise, $\dots$ shape neuronal networks' response to external world stimulations and what makes a linear response theory relevant in this context.  Understanding this is one contemporary task of theoretical neuroscience. 
The problem can be stated as follows. Assume that a neuronal network receives  a time-dependent stimulus $S(t)$ from time $t_0$ to time $t_1$. Even if the stimulus is applied to a subset of neurons in the network, its influence will eventually propagate to other neurons, directly or indirectly connected. This will result in a complex process where the effect of the stimulus is interwoven with neurons' dynamics.
 Yet, is there a way to disentangle, deconvolve, the effect of the stimulus from the "background" neuronal activity ?

A classical way to tackle this question is to consider the stimulation as a perturbation of a state of "spontaneous" activity. The response to the stimulation is then written as an expansion involving correlations, of higher and higher orders, computed with respect to the spontaneous activity. In the field of neuroscience this expansion is often called a Volterra expansion \cite{rieke-warland-etal:97}. The first term, where the response is proportional to the stimulus, is the linear response. 

Linear response theory is also at the core of non equilibrium statistical physics \cite{groot-mazur:84,gallavotti:14}. The response of a system, originally
at equilibrium, to a time-dependent perturbation is proportional to the stimulus, with proportionality coefficients
obtained via correlations functions of induced currents, where the correlations are computed with respect to the equilibrium distribution. These are Green-Kubo relations \cite{green:54,kubo:57}. The equilibrium state - which plays the role of the spontaneous state in the previous paragraph - is characterised, in statistical physics, by a Gibbs distribution. 

It is therefore interesting to try and connect neuronal networks models dynamics and non equilibrium statistical physics, although the link is not straightforward. Statistical physics certainly applies to, e.g., characterise ionic transfer at the level of neurons and synapses, but here we are addressing the question at another level. When modelling neuronal networks one uses simplified models where the microscopic activity (at the molecular level) has been averaged out to produce a dynamical system with reduced mesoscopic variables (local voltage of a neuron, conductances, concentration of neurotransmitters, ...). It is not evident \textit{a priori}  that the formalism of equilibrium and non equilibrium statistical physics can be used at this mesoscopic level, so as to characterise the neural response to stimuli. In particular, dynamics is clearly non Hamiltonian, not time reversible, dissipative. Therefore, in such an analogy, what should be the form of the "energy" in the Gibbs distribution? Do their exist the analogue of currents ? For which quantity ?
What is "transported" ? Although, there exist many approaches in neuroscience with a statistical physics flavour - the use of maximum entropy principle to analyse spike trains \cite{schneidman-berry-etal:06,shlens-field-etal:06,vasquez-cessac-etal:09}, the free energy principle from K. Friston et al \cite{friston-kilner-etal:06,friston:09,friston:10}, among many others - they are based on a formal analogy with statistical physics/thermodynamics principles, instead  of being derived from the collective neuronal dynamics, in a kinetic-like theory.

Thus, the main question we want to address in this paper is:\\

\begin{itemize}
\item \textit{Are there model-examples where the spontaneous spatio-temporal activity of a neuronal network can be characterised by a suitable form of Gibbs distribution, constructed from the neurons' dynamics, thereby allowing to derive a linear response theory where the linear response kernel is explicitly written in terms of the parameters governing the dynamics ?}
\end{itemize}

The resolution of this problem would also give hints to some pending questions in neuroscience:

\begin{itemize}

\item \textit{How does a stimulation applied to a subgroup of neurons in a population affect the dynamics of the whole network ?}

\item \textit{How to measure the influence of a neuron's stimulation on another neuron, especially if they are \textit{not} synaptically connected ?} This question is related to the notion of "effective" or "functional" connectivity between neurons (or groups of neurons) at the core of many researches in neuroscience \cite{friston:11b}. Several definitions of this connectivity can be given, not necessarily equivalent (e.g. based on pairwise correlations, causality, or mutual information \cite{battaglia-guyon-etal:17,cabral-kringelbach-etal:17}).

\item \textit{How does this connectivity relate to synaptic connectivity and non linear dynamics ? Is it related to a transported quantity, typically "information" ?}

\end{itemize}

In this paper, we address these points on the basis of two paradigmatic neuronal network models, the Amari-Wilson-Cowan model \cite{amari:72,wilson-cowan:72} and the conductance based Integrate and Fire model proposed by M. Rudolph and A. Destexhe in\cite{rudolph-destexhe:06,rudolph-destexhe:07}. The main idea here is to associate, to the spontaneous dynamics of the model, a natural notion of Gibbs distribution inherited from ergodic theory or stochastic processes and to derive the linear response from this.

This is a review paper, based on a lecture given in the LACONEU 2019 summer school in Valparaiso, \url{http://laconeu.cl/}. The main material has been published elsewhere although the presentation and discussion are original. It addresses both to the community of neuroscientists - interested in the applications of ideas and techniques coming from statistical physics and dynamical systems to neuroscience - and to experts in non equilibrium statistical physics interested in questions in neuroscience where their knowledge could be helpful.  Although based on a well established mathematical framework the results presented here, from a physicist point of view,  are non rigorous as we are most of the time at the border of theorems or far from these borders.

The paper is organised as follows. In section \ref{sec:bases} we give a brief introduction to neuronal modelling and non equilibrium statistical physics for non expert readers. Then, we develop two examples where a linear response theory can be derived for a neuronal network, with explicit relations between the parameters and equations defining the neurons dynamics and the linear response convolution kernel. This way, we give partial answers to the questions raised in this introduction. In section \ref{sec:LinRepFiring}, we present a former work with  J.A. Sepulchre where linear response can be lead relatively far using the fact that the network dynamics is chaotic \cite{cessac-sepulchre:04,cessac-sepulchre:06,cessac-sepulchre:07}. In section \ref{sec:LinRepSpiking}, we present an ongoing work with R. Cofr\'e \cite{cessac-cofre:19} where we study the derivation of a linear response in a spiking neural network model using the formalism of Markov chain and chains with complete connections\cite{onicescu-mihoc:35}. The last section is devoted to discussion.

\section{Bases}\label{sec:bases}

\subsection{Neuronal networks}\label{sec:NN}

We give here a brief summary of neuronal dynamics for the non familiar reader, so as to help him/her better understand the models presented later. For more details see \cite{gerstner-kistler:02,dayan-abbott:01,ermentrout-terman:10}.

\subsubsection{The dynamics of neuronal voltage} \label{sec:DynVoltage}

Neurons are cells able to use ionic transfers to change locally their membrane potential (voltage), that is, the difference between the local electric potential inside the cell and the local electric potential outside the cell. The fundamental equation controlling the time evolution of the voltage $V_k$ of neuron $k$ is based on local charge conservation:
\begin{equation}\label{eq:ChargeCons}
C_k \frac{dV_k}{dt} = i_{ion,k} + i_{syn,k} + i_{ext,k} + i_{noise,k},
\end{equation}
where $C_k$ is the neuron's membrane capacity. It relates the neuron's voltage variations to currents flowing through the neuron's membrane. The main currents are:\\

\paragraph{Ionic currents in the body of the neuron (soma, axon).} They are due to  fluxes of ions going through ionic channels in the membrane. These channels can open or close depending on the membrane voltage as well as other variables depending on the channel's type.
These currents are summarised by the term:
\begin{equation}\label{eq:i_ion}
i_{ion,k} = - \sum_{X} g_{k,X}\pare{.} \, (V_k - E_X),
\end{equation}
where the sum holds on ionic channels types, selective to ionic species (sodium, potassium, chloride, ...).
The quantity $g_{k,X}$ is the conductance of the channel of type $X$. It is roughly proportional to the density of open channels of type $X$ in the membrane.
It depends on hidden variables (activation, inactivation, $\dots$) summarised here as a dot ($.$) without further indication. $E_X$ is the reversal or Nernst potential of $X$ corresponding to the voltage $V_k$ at which the ionic current of type $X$ changes its direction.

Typical ionic currents are those generating action potentials, also called spikes. 
These are  fast (of order milliseconds), large increase in the membrane voltage (depolarisation) due the influx of positive ions (typically, sodium), followed by a fast decrease (re-polarisation) due to the efflux of potassium, then by a refractory period during which the neuron cannot emit a spike any more.
If the shape of the spike is essentially invariant for a given neuron, the sequence of spikes it emits may vary significantly in timing.   The term $i_{ion,k}$ can contain many others ionic current types, regulating the neuron's activity, in particular, spikes timing (see \cite{ermentrout-terman:10} for a nice mathematical and biophysical presentation).

\paragraph{Synaptic currents.}  An increase (depolarisation) in the voltage of the pre-synaptic neuron $j$  induces a release of neurotransmitters which diffuse to the post-synaptic neuron $k$ and bind to specific receptors with different possible mechanisms. This triggers the opening of ionic channels generating a local (at the level of the synapse) synaptic current, $i_{syn,k}$ that takes a similar form as \eqref{eq:i_ion}:
\begin{equation}\label{eq:i_syn}
i_{syn,k} = - \sum_{j} g_{kj}\pare{.} \, (V_k - E_{kj}),
\end{equation}
although the mechanisms regulating the synaptic conductance $g_{kj}$ from $j$ to $k$ are of a different nature. (See \cite{destexhe-mainen-etal:94b} for a clear and synthetic presentation of neurotransmitter-receptors modelling).
The sum holds on pre-synaptic neurons connected to $k$. $E_{kj}$ is the reversal potential of the ionic species triggering the current from $j$ to $k$. 

\paragraph{External current.}
Neurons can be excited by external stimuli (typically an external current imposed by an electrode). This corresponds to the current $i_{ext,k}$ in \eqref{eq:ChargeCons}. In general, it depends explicitly on time. In the paper, $i_{ext,k}$ will design a stimulus exogenous to the network, without specifying how it is injected to the neurons (it could be e.g. a synaptic current coming from neurons of another network).

\paragraph{"Noise" current.}
Finally, $i_{noise,k}$ is a stochastic term that mimics "noise" in neurons' dynamics (thermal noise inducing a probabilistic ionic channel activation, diffusion of neurotransmitters, $\dots$). In general, $i_{noise,k}$ is modelled as a white noise.


Note that, on biophysical grounds equation  \eqref{eq:ChargeCons} holds for a small piece of the neuron's membrane. In this paper, though, we will consider neurons as points (no spatial structure), so that eq. \eqref{eq:ChargeCons} holds for the whole neuron. 

\subsubsection{Collective response} \label{sec:ColRep}

The collective dynamics of neurons \eqref{eq:ChargeCons} in the absence of stimulation ($i_{ext,k}(t)=0$) is called \textit{spontaneous}. It results from the intrinsic non linear dynamics of neurons (the term $i_{ion,k}$) as well as from neurons' interactions (the term $i_{syn,k}$). It can therefore be quite complex (bursting \cite{izhikevich:00} , chaotic\cite{aihara-takabe-etal:90,preissl-lutzenberger-etal:96,korn-faure:03}, generating waves \cite{bressloff:14}, $\dots$). In addition, the noise term 
introduces some degree of stochastic. Thus, in general, one is not attempting to analyse
the individual trajectories of the dynamical system  \eqref{eq:ChargeCons}, but instead, one is studying statistical properties (e.g. spike rates or spikes correlations). It is a reasonable assumption, used in all the model we know, to consider that statistics of the spontaneous activity is stationary (time-translation invariant). This means that neurons' spike rates in spontaneous activity are constant, or that the pairwise spike correlation between $2$ neurons only depends on the time interval between the spikes emitted by these neurons.

When the neuronal network is submitted to an external influence (the term $i_{ext,k}$ in \eqref{eq:ChargeCons}), the collective dynamics is called \textit{stimulus evoked} or \textit{stimulus dependent}.
As the stimulation usually depends explicitly on time, the stationarity assumption, \textit{stricto-sensu} does not hold. Yet, many theoretical tools are grounded on a stationarity assumption: especially all methods based on entropy  (maximum entropy, mutual information). 
Other approaches, as the one presented in this paper, are not mathematically constrained by this hypothesis. 
 The alternative strategy used here considers that the stimulus has a small amplitude, so that the neuronal network responds proportionally to the stimulus. In other words, the difference between spontaneous activity statistics and evoked statistics is proportional to the stimulus amplitude. In the context of linear response theory the proportionality coefficient is computed from correlations functions  in spontaneous activity. 
 
\subsubsection{Simplified models} \label{sec:SimpModels}

Equation \eqref{eq:ChargeCons} hides a large number of additional differential equations ruling conductances, calcium dynamics, synaptic activity and so on, which are just impossible to study mathematically in full generality. They are also hard to simulate, not only because of the large number of variables and equations, but also, and mainly, because of the large number of parameters entering in the biophysics, which have to be determined from experiments.
Modellers are therefore using simplified versions of \eqref{eq:ChargeCons} focusing on some specific aspects and questions, as we will do in this paper.

As a salient feature of most neurons in the nervous system (although not all of them) is to produce spikes, a standard modelling approach is to characterise the activity of a neuron or of a neuronal network by focusing  on spikes timing. Neural activity is then represented by spike trains, or firing rates - the number of spikes emitted per second by a neuron. A firing rate model is presented in section \ref{sec:LinRepFiring}. A spiking model is presented in section \ref{sec:LinRepSpiking}.

\subsection{Linear response in statistical physics}\label{sec:StatPhys}

We give a brief summary of non equilibrium statistical physics (with a physicists point of view) for non familiar readers. We adopt a classical description (see \cite{le-bellac-mortessagne-etal:04} for a didactic, yet wide, introduction to the subject) disregarding, for the moment, more general and synthetic recent approaches \cite{baiesi-maes:13,basu-maes:15} (see discussion section).

We consider a system characterised by a microstate $\omega$ (a point in the phase space, a spin configuration, $\dots$). For simplicity we consider that $\omega$ takes a countable number of values. 

When the system is at equilibrium the probability to observe the microstate $\omega$ is:
\begin{equation}\label{eq:GibbsDef}
\Prob{\omega}=\frac{1}{Z} \, e^{-\frac{H\pare{\omega}}{k_B T}}
\end{equation}
where $Z=\sum_{\omega} e^{-\frac{H\pare{\omega}}{k_B T}}$ is called partition function, with $k_B$, the Boltzmann constant and $T$ the temperature in Kelvin. The function of $\omega$:
\begin{equation} \label{eq:HStatPhys}
H\pare{\omega} =  \sum_{\alpha} \lambda_\alpha X_\alpha\pare{\omega},
\end{equation}
is called the energy of the microstate $\omega$. The functions $X_\alpha$ are extensive quantities (proportional to the number of particles) such as energy, electric charge, volume, number of particles, magnetic field, $\dots$ 
The conjugated parameters $\lambda_\alpha$ 
correspond to intensive quantities (not proportional to the number of particles), like temperature, electric potential, pressure, chemical potential, $\dots$. In general they depend on the location in the physical space (e.g. the temperature depends on the position in a fluid). At equilibrium they are uniform in space though. 

The form of $H$, i.e. the choice of the $\lambda_\alpha$ and $X_\alpha$ is constrained by the physical properties of the system. It is also constrained by boundary conditions. 
In standard statistical physics courses, the Gibbs distribution form \eqref{eq:GibbsDef} is obtained as a consequence of a principle, the Maximum Entropy Principle \cite{jaynes:57}:
maximising the statistical entropy under the constraint that the average value of $H$ is fixed. The statistical entropy is proportional to the Shannon entropy (up the, fundamental, Boltzmann constant), making a deep link between thermodynamics and information theory. More general definitions of Gibbs distributions exist though, not constrained by entropy, and constructed from dynamics. We see two examples in this paper. In this setting, maximising entropy is a consequence of large deviations theory \cite{dembo-zeitouni:98}. 

Equilibrium statistical physics allows to establish macroscopic laws from first principles, These laws summarise a complex, non linear dynamics, with a large number of particles, in a few macroscopic variables related by a few equations \cite{le-bellac-mortessagne-etal:04}. A well known example is the law of ideal gas, $PV=nRT$. A natural question is whether spontaneous neuronal dynamics could obey similar laws.\\

A non equilibrium situation arises when the $\lambda_\alpha$s are not uniform in space, generating gradients ("thermodynamic forces"), $\vec{\nabla} \lambda_\alpha$ (temperature gradient, electric potential gradient ...). These gradients result in currents density $\vec{j}_\alpha$ of $X_\alpha$ (e.g. a temperature gradient induces a heat current). In Onsager theory, currents density are functions of gradients: 
$$ 
\vec{j}_\alpha= \vec{\cF}_\alpha(\vec{\nabla} \lambda_1, \dots, \vec{\nabla} \lambda_\beta, \dots ).
$$
If gradients are small and if $\vec{\cF}_\alpha$ is differentiable:
$$ \vec{j}_\alpha \sim \sum_{\beta } L_{\alpha \beta} \vec{\nabla} \lambda_\beta + \dots,$$
where the coefficients $L_{\alpha \beta}$ are called Onsager coefficients \cite{onsager:31}.

Typical examples are the Ohm's law where the electric current density 
$\vec{j}_{el} = -\sigma_E \vec{\nabla} V$, is proportional to the gradient of electric potential with a factor $\sigma_E$, the electric conductivity, or the Fourier's law
 $\vec{j}_{Q} = -\lambda \vec{\nabla} T$ where the heat flux is proportional to the temperature gradient. 

Assuming the gradients are small enough so that the system can be divided into mesoscopic cells at equilibrium (quasi static approximation) the Onsager coefficients can be derived as correlation functions 
computed with respect to the equilibrium distribution. This constitutes the Green-Kubo relations \cite{green:54,kubo:57}:
\begin{equation}\label{eq:GreenKubo}
L_{\alpha\beta} \propto \int_0^{+\infty} \brk{\vec{j}_\alpha(0).\vec{j}_\beta(s)}_{eq} \, ds,
\end{equation}
where $\brk{}_{eq}$ denotes the average with respect to the Gibbs equilibrium probability \eqref{eq:GibbsDef}. We assume here that $\brk{\vec{j}_\alpha(0)}_{eq}=\brk{\vec{j}_\beta(s)}_{eq}=0$ so that \eqref{eq:GreenKubo} is the time integral of correlation of currents, where correlations are computed at equilibrium.

An important relation, which allows to derive Onsager coefficients from dynamics in several examples \cite{gallavotti:96} is the entropy production, $\sigma \equiv  \frac{dS}{dt}$. It is given, to the first order \cite{maes-h-van-wieren:06}, in terms of  the Onsager coefficients by:

\begin{equation}\label{eq:EntropyProd}
\sigma =\sum_\alpha \vec{\nabla} \lambda_\alpha.\vec{j}_\alpha = \sum_{\alpha,\beta}  \vec{\nabla} \lambda_\alpha  L_{\alpha \beta} \vec{\nabla} \lambda_\beta.
\end{equation}

The interesting point is that the Green Kubo relations can be obtained, in some cases, from the microscopic dynamics ruling the evolution of the microstate, using different techniques: this has been done in dynamical systems and ergodic theory \cite{gallavotti-cohen:95,gallavotti-cohen:95b,gallavotti:96,ruelle:99}, or stochastic processes and Markov chains (See \cite{kaiser-jack-etal:18} and references therein). A recent method, synthetising previous approaches has also been proposed by C. Maes and co-workers \cite{baiesi-maes:13,basu-maes:15}. The application of these formalisms allows to construct a linear response theory in neuronal models, from the equations ruling neurons' dynamics, as we now show.

%
%
%
%
%


\section{From firing rate neurons dynamics to linear response}\label{sec:LinRepFiring}

\subsection{The Amari-Wilson-Cowan model} \label{sec:AWC}

As a first example of linear response in neural network we consider a canonical model of neuronal dynamics, the Amari-Wilson-Cowan model \cite{amari:72,wilson-cowan:72,wilson-cowan:73}. It consists of a set of $N$ neurons, $i=1 \dots N$, with membrane voltage $V_i$, whose dynamics is given by the dynamical system:
\begin{equation}\label{eq:AWC}
\frac{d V_i}{d t} = - V_i + \sum_{j=1}^N J_{ij} f(V_j(t)) + \epsilon S_i(t); \quad i=1 \dots N. 
\end{equation}
This equation can be derived from equation  \eqref{eq:ChargeCons} up to several approximations explained e.g. in \cite{ermentrout:98,faugeras-touboul-etal:09}.
Note that the decay (leak) term $- V_i$ has a more general form $-\frac{1}{\tau} (V_i-V_L)$ where $V_L$ is a reversal potential and $\tau$ is a characteristic time. It is easy to get to the form \eqref{eq:AWC} by rescaling time and voltages.

We will also consider the discrete time version of \eqref{eq:AWC}:
\begin{equation}\label{eq:AWC_Discrete}
V_i(t+1)= \sum_{j=1}^N J_{ij} f(V_j(t)) + \epsilon S_i(t); \quad i=1 \dots N.
\end{equation}
This form is more convenient to understand the mechanisms that shape the linear response, namely the interplay between neurons interactions and non linear dynamics, because one can follow the effect of a stimulus time-step by time-step. For this reason we will mainly stick at the discrete time dynamics, except in the next subsection (Contractive regime), where the derivation of the linear response and its consequences are more straightforward and familiar to readers in the continuous time case.

Neurons are coupled via synaptic weights $J_{ij}$  characterising the strength of interaction from the pre-synaptic neuron $j$ to the post-synaptic neuron $i$. This defines an oriented graph, i.e. $J_{ij} \neq J_{ji}$ in general, in contrast to physics where interactions are symmetric. This graph is also signed: when $J_{ij} >0$ the interaction (synapse) is  excitatory, when $J_{ij} < 0$, it is inhibitory. 

\begin{figure}
\center
\includegraphics[width=8cm, height=4cm]{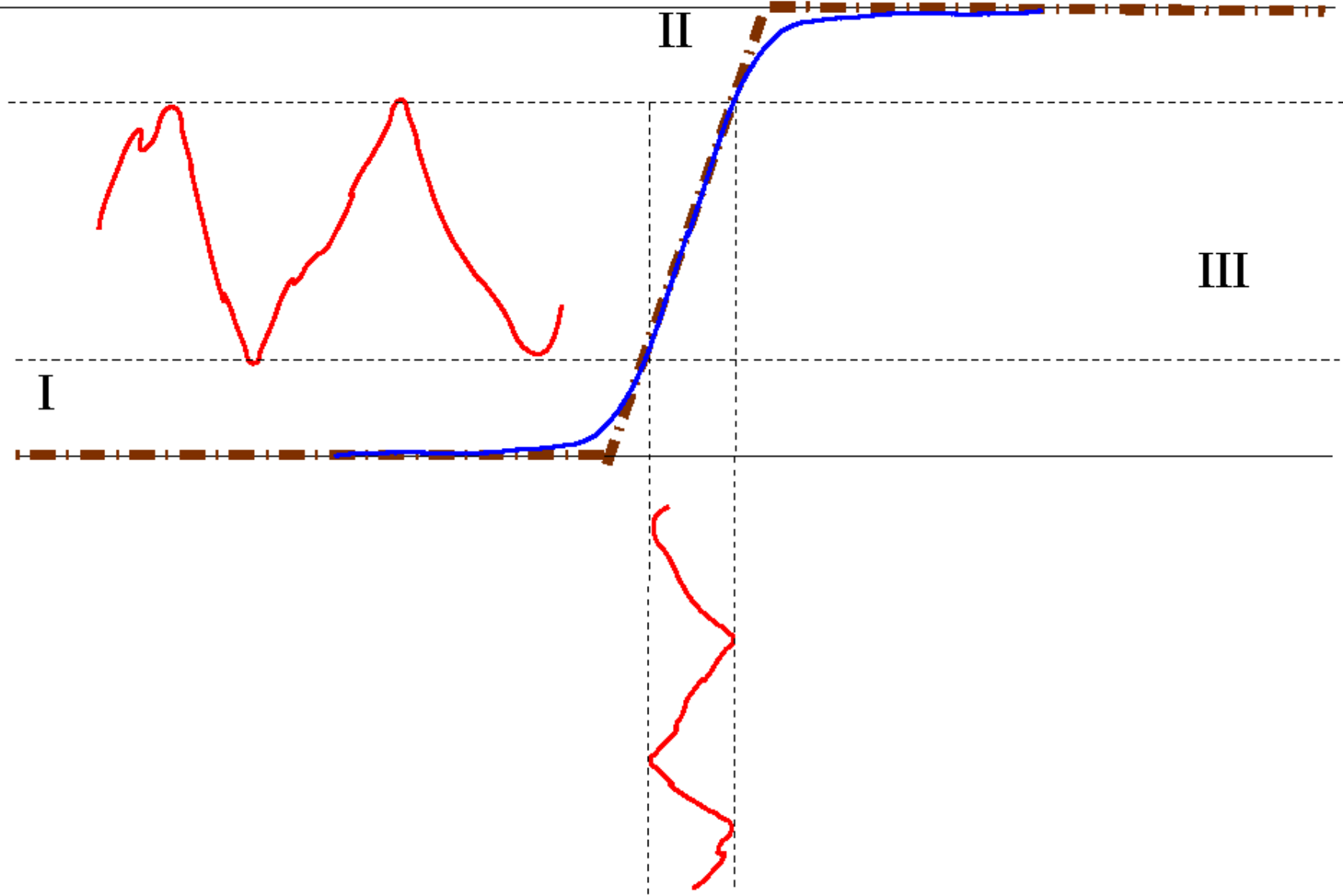}

\vspace{0.5cm}

\includegraphics[width=8cm, height=4cm]{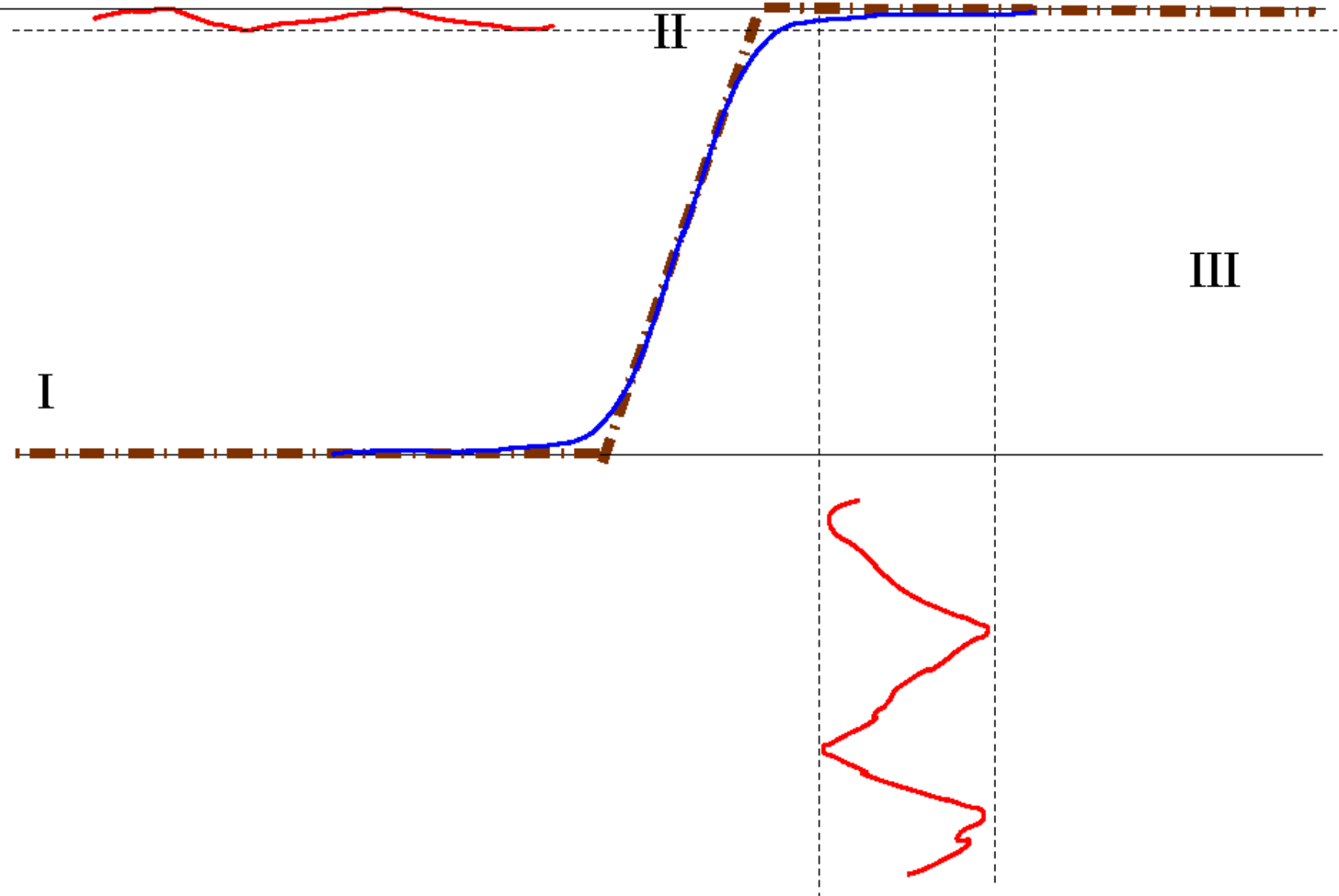}
\caption{\label{fig:ExpCont} The sigmoidal shape of the function $f$ and its effects on voltage fluctuations. Top. When the neuron's voltage fluctuates around the inflection point of the sigmoid, and if the gain $g$ is large enough fluctuations are amplified. Bottom. When the neuron's voltage fluctuates in the flat parts of the sigmoid (here, the saturated region) fluctuations are damped. Dashed red lines correspond to a piecewise linear approximation of the sigmoid, allowing to delimit regions $I$, $II$, $III$.}
\end{figure}

In this model, the pre-synaptic neuron $j$ influences the post synaptic neuron $i$ via its firing rate (probability to emit a spike in a small time interval) which is a function $f(V_j)$ of the pre-synaptic neuron voltage. Here, $f$ is a non linear, sigmoid function as depicted in Fig. \ref{fig:ExpCont}. A typical form for $f$ is:
\begin{equation} \label{eq:Sigmoid}
f(x)=\frac{1}{2}\pare{1+\tanh(g\, x)}.
\end{equation}
The sigmoidal shape has a deep biological importance.
Indeed, one can distinguish $3$ rough regions (Fig. \ref{fig:ExpCont}). In region I (low voltage), the neuron does not emit spikes. In region II, $f(V)$ is roughly linear. In region III (high voltage), the firing rate reaches a plateau, fixed by the refractory period. 

The parameter $g$ in \eqref{eq:Sigmoid}, either called  "gain" or "non linearity",  is of up most importance. On biophysical grounds, it characterises the sensitivity of the neuron's firing rate to fluctuations of its voltage. Consider indeed  Fig. \ref{fig:ExpCont}, top. When $g$ is larger than $1$ the fluctuations are amplified by $f$ in region $II$.
In contrast, in region $I$ and $III$ they are damped. This remark, made at the level of single neuron, has a deep importance when interpreting the linear response of a network governed by eq. \eqref{eq:AWC} or \eqref{eq:AWC_Discrete}. On dynamical grounds this effect controls the local expansion / contraction in the phase space, as developed below. 

Finally, in eq. \eqref{eq:AWC}, \eqref{eq:AWC_Discrete}, $S_i(t)$ is an "external stimulus". Here, it depends only on time but the analysis made below affords a situation
where $S_i$ depends also on the neuron's voltage.

We introduce the state vector $\vV=\vect{V_i}_{i=1}^N$,
the stimulus vector $\vS=\vect{S_i}_{i=1}^N$ and the matrix of synaptic weights $\cJ=\vect{J_{ij}}_{i,j=1}^N$.
With a slight abuse of notations we write $f(\vV)$ for the vector $\vect{f(V_i)}_{i=1}^N$. Then, we may rewrite \eqref{eq:AWC} in vector form:
\begin{equation}\label{eq:AWC_vect}
\frac{d \vV}{d t} = \underbrace{-  \vV + \cJ. f(\vV)}_{\vF(\vV)} + \epsilon  \vS(t),
\end{equation}
whereas \eqref{eq:AWC_Discrete} becomes:
\begin{equation}\label{eq:AWC_Discrete_vect}
\vV(t+1)= \underbrace{\cJ.f(\vV(t))}_{\vG(\vV(t))} + \epsilon \vS(t).
\end{equation}

Here, we don't make any hypothesis on the matrix of weights $\cJ$, except that it's entries are bounded in absolute value. This implies that the spontaneous dynamics ($\epsilon =0 $) can be restricted to a compact set $\cM$. 

We remark that the Jacobian matrix of \eqref{eq:AWC_vect}, $D\vF_{\vV}$ have the form:
\begin{equation}\label{eq:DF}
D\vF_{\vV}=- \cI + \cJ.\cD(\vV),
\end{equation}
where $\cI$ is the $N \times N$ identity matrix and $\cD(\vV)$ is the diagonal matrix:
\begin{equation}\label{eq:diagD}
\cD(\vV) = diag \pare{f'(V_i), \, i=1 \dots N}.
\end{equation}
Likewise, the Jacobian matrix of \eqref{eq:AWC_Discrete_vect}, $D\vG_{\vV}$, reads:
\begin{equation}\label{eq:DG}
D\vG_{\vV}=\cJ.\cD(\vV).
\end{equation}

\subsection{Contractive regime}\label{sec:Contractive}

\subsubsection{Dynamics}\label{sec:DynCont}

In order to illustrate the main ideas of this section we start by considering a specific regime of the Amari-Wilson-Cowan model, the contractive regime \cite{matsuoka:92,cessac:94}. This regime 
occurs when neurons have a very low gain $g$. In this case, the dynamical systems \eqref{eq:AWC_vect} and \eqref{eq:AWC_Discrete_vect} have a unique fixed point attracting all trajectories (absolutely stability). On mathematical grounds, this is the starting point for the analysis of the bifurcations cascade occurring in the model when increasing $g$. On neuronal grounds, this is a low reactivity regime where, after a sufficiently long time and whatever the initial condition, neurons fire with a constant rate in spontaneous activity. Although this regime is somewhat trivial it brings nevertheless several interesting hints of linear response in more complex situations.

The existence of a contractive regime is deduced from the following argument. If $g=0$, $\cD(\vV)=0$ and all eigenvalues of $D\vF_{\vV}$ are equal to $-1$ (resp. all eigenvalues of $D\vG_{\vV}$ vanish). By continuity, for $g$ sufficiently small, all the eigenvalues of $D\vF_{\vV}$ have a strictly negative real part, $\forall \vV \in \cM$, (resp. all the eigenvalues of $D\vG_{\vV}$ have a modulus strictly smaller than $1$). From this, one can establish that there is a $g$ value, $g_{as}(\cJ)$,  depending on $\cJ$, such that, for $g < g_{as}(\cJ)$,  the spontaneous dynamical system ($\epsilon =0$) has a unique fixed point $\vVa$ attracting all trajectories in $\cM$. When $g$ further increases, one gets out of this regime by a cascade of bifurcations leading to a more and more complex dynamics, discussed in section \ref{sec:Chaos}.

We consider now the perturbed neural network ($\epsilon >0$) in this contractive regime. We first make the computation for the continuous time system as it is straightforward and familiar to most readers. We look for solutions of the form $\vV=\vVa+\vxi$.
This is a standard linear stability analysis.
We have:
$$
\frac{d \vxi}{dt} = D\vF_{\vVa}.\vxi + \epsilon \vS(t) + O\pare{\epsilon^2},
$$
with solution:
%
$$
\vxi(t) = e^{D\vF_{\vVa}(t-t_0)}.\vxi(t_0) + \epsilon \int_{t_0}^t e^{D\vF_{\vVa}(t-s)}.\vS(s)ds + O\pare{\epsilon^2},
$$
%
where $t_0$ is the initial time where we start to apply the stimulus. This solution contains a transient term, dependent on the initial condition, and a stimulus dependent term. We now consider the steady state regime corresponding to $t_0 \to -\infty$: the stimulus was applied in a very distant time in the past, quite longer than the longest characteristic time of the dynamics. In this limit: 
\begin{equation}\label{eq:SolLinStablePermanent}
\vxi(t) \sim \epsilon \, \int_{-\infty}^t e^{D\vF_{\vVa}(t-s)}.\vS(s)ds = \epsilon \, \bra{\chi \ast S}(t),
\end{equation}
where the integral converges since all eigenvalues have negative real part. We have introduced the matrix:
\begin{equation}\label{eq:Chi_Stable_Cont}
\chi(t)=e^{D\vF_{\vVa} t}.
\end{equation}
Equation \eqref{eq:SolLinStablePermanent} is a first example of a linear response formula, where the deviation $\xi$ from the solution of spontaneous dynamics (here, the fixed point $\vVa$), is proportional to the stimulus, and expressed by a convolution with the linear response kernel \eqref{eq:Chi_Stable_Cont}. Thus, the linear response kernel $K$ discussed in the introduction reads $K=\epsilon \chi$.

The same result holds \textit{mutatis mutandis} for the discrete time dynamical system \eqref{eq:AWC_Discrete_vect} with a discrete time convolution:
\begin{equation} \label{eq:SolLinStablePermanentDiscrete}
\xi(t)= \epsilon \, \sum_{\tau=-\infty}^{t-1} DG^{t-\tau-1}_{\vVa}.\vS(\tau)  + O(\epsilon^2)
\end{equation}
where :
\begin{equation}\label{eq:Chi_Stable_Disc}
\chi(t)=
D\vG^{t}_{\vVa},
\end{equation}
the $t$-th iterate of Jacobian matrix $D\vG$ at $\vVa$. 
We come back to the derivation of \eqref{eq:SolLinStablePermanentDiscrete}, for a more general case, in section \ref{sec:Chaos}. The series \eqref{eq:SolLinStablePermanentDiscrete} converges because, in the contractive regime, the eigenvalues of $D\vG_{\vVa}$ have a modulus strictly lower than $1$.

\subsubsection{Susceptibility and resonances}  \label{sec:Susc} 

The Fourier transform of \eqref{eq:SolLinStablePermanent}  is:
\begin{equation}\label{eq:LinRepFourier} 
\hat{\xi}(\omega)=\epsilon \hat{\chi}(\omega).\hat{S}(\omega),
\end{equation}
where $\omega$ is a real frequency.   $\hat{\chi}(\omega)$, called the complex susceptibility, characterises the response to a harmonic stimulus with frequency $\omega$,
with an amplification factor (the modulus $|\hat{\chi}(\omega)|$) and a phase factor (the argument of $\hat{\chi}(\omega)$). $\hat{\chi}(\omega)$  can be analytically continued in the complex plane (complex frequencies, $\omega = \omega_r + i \omega_i$). Its  explicit computation  is straightforward. We give it first in the continuous time case.  We note $\lambda_k = \lambda_{k,r} + i \lambda_{k,i}$ the eigenvalues of $D\vF_{\vVa}$. Note that, as the matrix $\cJ$ is not symmetric, the eigenvalues $\lambda_k$ are complex in general. Denoting $P$ the transformation matrix diagonalising $D\vF_{\vVa}$ we have $\hat{\chi}(\omega)=P.\hat{\chi}'(\omega).P^{-1}$
where $\hat{\chi}'_k(\omega)=\hat{\chi}'_{k,r}(\omega) \, + \, i \, \hat{\chi}'_{k,i}(\omega)$ is the diagonal matrix with entries:
\begin{equation}\label{eq:chiDiag}
\begin{array}{lll}
\hat{\chi}'_{k,r}(\omega) &=& - \frac{\lambda_{k,r}+\omega_i}{\pare{\lambda_{k,r}+\omega_i}^2+\pare{\lambda_{k,i}-\omega_r}^2},\\
\hat{\chi}'_{k,r}(\omega) &=& \frac{\lambda_{k,i}-\omega_r}{\pare{\lambda_{k,r}+\omega_i}^2+\pare{\lambda_{k,i}-\omega_r}^2}.
\end{array}
\end{equation}

\begin{figure}
\center
\includegraphics[width=8cm, height=6cm]{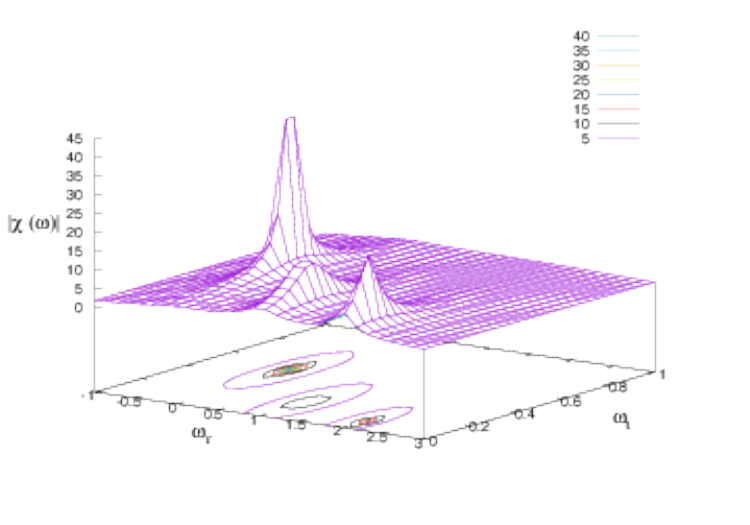}

\vspace{0.5cm}

\includegraphics[width=7cm, height=5cm]{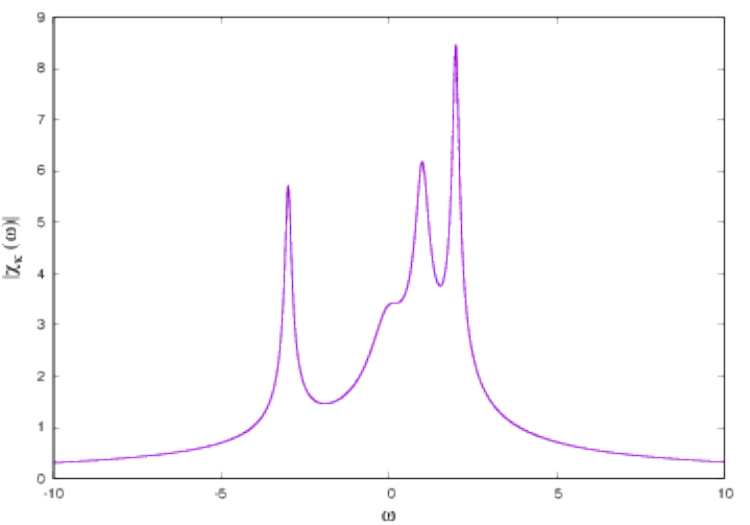}
\caption{\label{fig:Resonances_FP} Schematic plot of the resonances. Up. The modulus $\abs{\hat{\chi}(\omega)}$ for complex frequencies. Bottom. $\abs{\hat{\chi}(\omega)}$ projected on the real axis.}
\end{figure}

The salient remark is that this expression has poles in the complex plane, located at $\omega_r=\lambda_{k,i}$, 
$\omega_i=-\lambda_{k,r}$, $k=1 \dots N$. 
On the real axis (real frequencies) the trace of these poles gives peaks (resonances) where the response to harmonic perturbation is maximal. The peaks are located at $\omega_r=\lambda_{k,i}$ while their width depends on $\omega_i=-\lambda_{k,r}$. The situation is sketched in Fig. \ref{fig:Resonances_FP}, where we have plotted $\abs{\hat{\chi}(\omega)}$. We have only shown $4$ poles for the legibility of the figure (2 poles are close to each other, as better seen on the real axis projection, bottom figure). 

A similar computation can  be done in the discrete time case. If $\lambda_k$ denote now the eigenvalues of $D\vG_{\vVa}$ the diagonal susceptibility $\hat{\chi}'(\omega)$ has entries $\hat{\chi}'_k(\omega)=\frac{1}{1-\lambda_k \, e^{i \omega}}$. Resonances are therefore now given by $\lambda_k = e^{-i \omega}$.

\subsubsection{The neural interpretation of resonances}  \label{sec:Resonances} 

The existence of these resonances means the following.  Exciting a neuron, or a group of neurons, with a harmonic perturbation, there are frequencies 
where the response to the perturbation is maximal. 
Especially, groups of neurons synchronise for a given resonance frequency. These groups depend on the corresponding eigenvector of the Jacobian matrix. 

Let us understand now how these resonances relate to neurons' interactions and non linear dynamics.
This is actually simpler for the discrete time dynamical system \eqref{eq:AWC_Discrete}, mainly because we are going to study the propagation, step by step, of a signal along the network edges.

Assume therefore that we are injecting a periodic stimulus $S$, with frequency $\omega$, at only one neuron, say $j$. How does this stimulation affect the other neurons in the network ? In the contractive regime the answer is relatively simple because neurons, in spontaneous activity, have a constant voltage (fixed point $\vVa$). When the stimulus is injected at neuron $j$, its voltage $V_j$ oscillates periodically around this equilibrium value $V_j^\ast$. These oscillations are then synaptically transmitted to its post synaptic neighbours. If $\epsilon$ is small enough, the synaptic action of $j$ to neuron $k_1$ is $\epsilon \, J_{k_1j} \, f'(V_j^\ast) \, S_j(t)$ to order $1$ in $\epsilon$. Thus, it is proportional to the synaptic weight, and to the derivative of $f$ at $V_j^\ast$. 

We find the effect illustrated in fig. \ref{fig:ExpCont}: depending on which region of the sigmoid neuron $j$'s potential is, the fluctuation induced by the stimulus are either non linearly contracted (in region I, III) or linearly multiplied by $g$ in region II (where $g$ is smaller than $1$ in the contractive regime, in contrast to fig. \ref{fig:ExpCont} top). 

\begin{figure}
\center
\includegraphics[width=6cm, height=5cm]{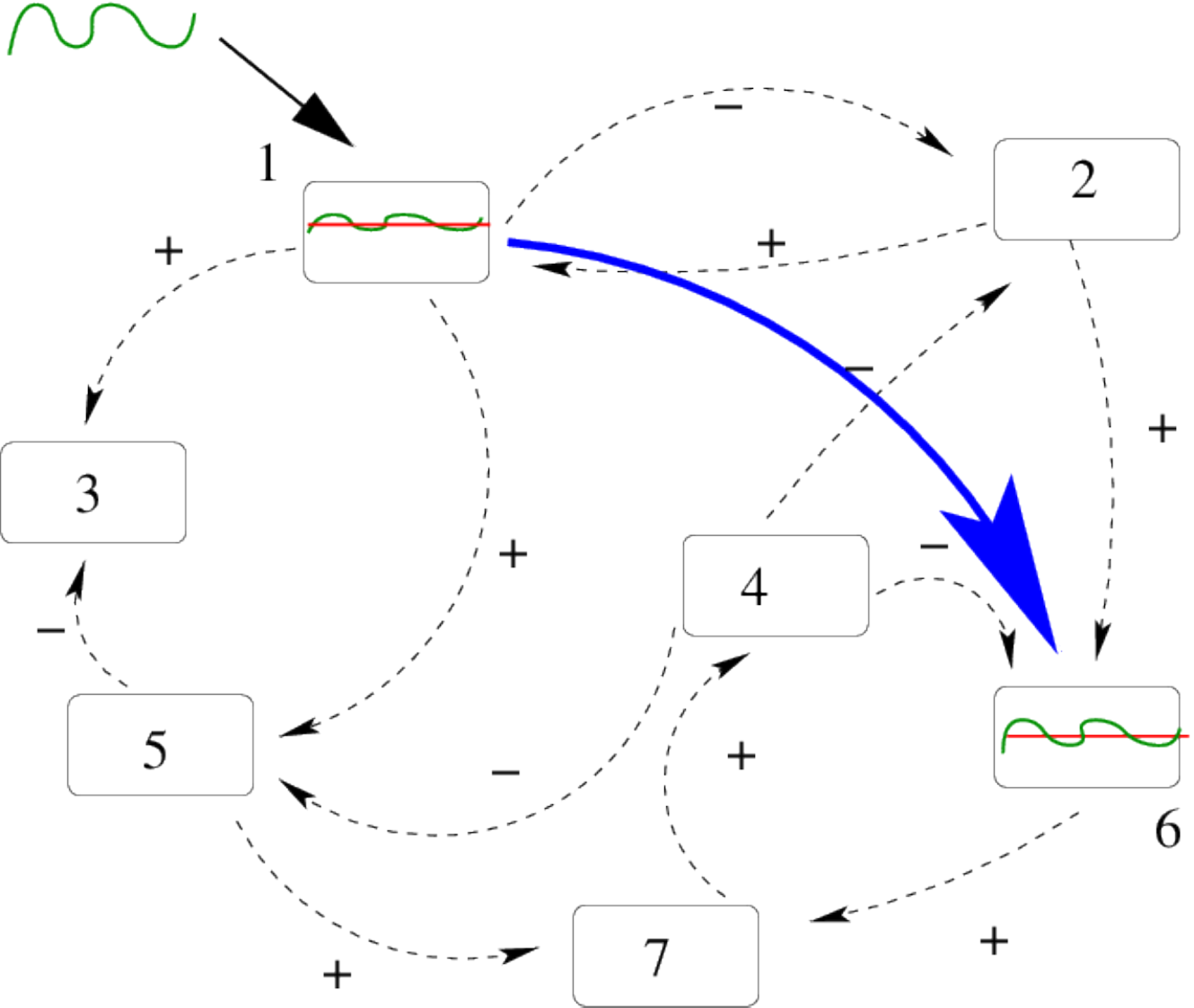}

\vspace{0.5cm}

\includegraphics[width=6cm, height=5cm]{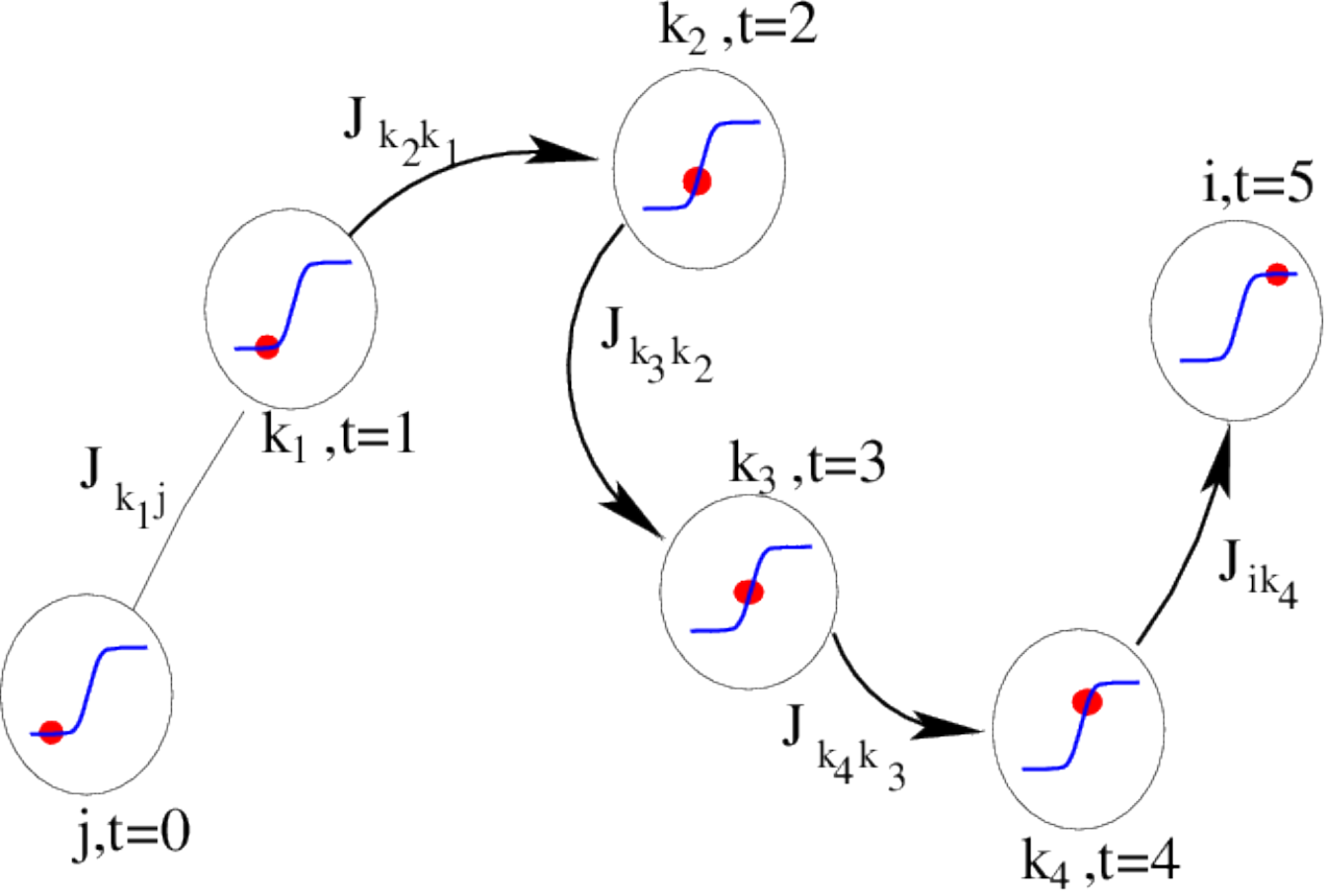}
\caption{Propagation of a stimulation throught the network. Top. A stimulus (green trace) is applied to neuron $1$, superimposed upon the rest activity (red trace). This stimulation propagates, through the network, up to neuron $6$ inducing a response (green trace in the box $6$). Thus, the stimulation of neuron $1$ influences neuron $6$ even if there is no direct pathway between them. This is summarised by the blue arrow. This effect results from the summation of the stimulus influences propagating through network pathways, depending on neurons' voltage
along those pathways. This is represented in the bottom figure. The state of the neuron (fixed point) is represented in red on the graph of the sigmoid function. The stimulus response is proportional to the derivative of the sigmoid at the red point. 
   \label{fig:Circuits}}
\end{figure}

More generally, we see that the first-order effect of the stimulus applied at $j$, on a neuron $i$, connected to $j$ via a synaptic path $j \to k_1 \to k_2 \to k_3 \to k_4 \to i$, as in fig. \ref{fig:Circuits}, is proportional to the stimulus $\epsilon S_j(t)$ with a proportionality coefficient $\prod_{l=1}^5 J_{k_lk_{l-1}} f'(V_{k_{l-1}}^\ast)$, where we set $k_0=j$ and $k_5=i$. 

Thus, the effect of a stimulus applied to $j$ at time $0$, on neuron $i$, $\sigma$ time step later, is proportional to:
\begin{equation}\label{eq:chi_Contract}
\chi_{i,j}(\sigma) =
\sum_{\gamma_{ij}(\sigma)}
        \prod_{l=1}^{\sigma}J_{k_l k_{l-1}}
\prod_{l=1}^{\sigma}f'(V_{k_{l-1}}^\ast),
\end{equation}
where the sum holds on all paths $\gamma_{ij}(\sigma)$ connecting $j$ to $i$ in $\sigma$ time steps.
This coefficient is nothing but the entry $ij$ of $D\vG^\sigma_{\vVa}$, the Jacobian matrix of the $\sigma$-th iterate of the mapping $\vG_{\vVa}$. This gives an interesting, network oriented, interpretation of  equation \eqref{eq:Chi_Stable_Disc} where one can explicitly compute the linear response convolution kernel with an explicit dependence in the parameters determining the network dynamics.

When applying a periodic stimulus to neuron $j$, the effect propagates through the network edges, with a weight proportional to the synaptic weight and to the derivative $f'$ at the corresponding node (neuron), as in Fig. \ref{fig:Circuits}. The effects of all these paths sum up at neuron $i$, generating contributions that can add up or interfere. This depends on the matrix $\cJ$ and on the neurons' rest state $\vVa$. It also depends on the frequency $\omega$. At resonance frequencies the effects of the stimulus cumulate in an optimal way, generating a maximal response. 

We have seen that these resonances are given by the eigenvalues of the Jacobian matrix $D\vG_{\vVa}$. Interestingly, eigenvalues of a matrix can be expressed in cyclic expansions (using trace formula and $\zeta$ functions \cite{ruelle:76b,parry-pollicott:90,gaspard:05}, see also \url{http://chaosbook.org/}), in terms of closed loops in the connectivity circuit defined by the matrix $D\vG_{\vVa}$. Resonances correspond therefore to constructive interferences along these closed loops. This establishes a nice correspondence between the dynamical response to a stimulus, the network topology and the non linear dynamics. 

\subsection{Chaotic dynamics}\label{sec:Chaos}

\subsubsection{Beyond the contractive regime}\label{sec:BeyondCont}

We now consider the Amari-Wilson-Cowan model outside the contractive regime, when $g$ is larger than $g_{as}(\cJ)$. There, strong non linear effects take place. Increasing $g$ beyond the contractive regime results in general in codimension $1$ local or global bifurcations \cite{doyon-cessac-etal:93,cessac-doyon-etal:94}, the most common being the destabilisation of the fixed point, $\vVa$, by a Hopf bifurcation, giving rise to periodic oscillations; or the appearance of other fixed points by saddle node bifurcation (or pitchfork when $f(x)$ has the symmetry $f(x)=-f(-x)$).

Here, we will consider a situation where dynamics become chaotic when $g$ is large enough. On neuronal grounds, chaotic network have deserved a lot of attention since the 80s. Especially, the neurobiologist W. Freeman \cite{freeman-yao-etal:88} argued that chaos in the brain may provide the brain vital flexibility while allowing it to visit many states in quick succession.  The underlying idea is that chaotic dynamics constitute an infinite reservoir of information, that a neuronal network can access e.g. via a proper training (typically Hebbian learning or spike time dependent plasticity). This idea is at the core of many applications of neural networks such as echo state machines, liquid state machines, deep learning. 
The Amari Wilson Cowan model is a paradigmatic example exhibiting chaos where one can illustrate the computational properties of neuronal networks at the edge of chaos \cite{bertschinger-natschlager:04,siri-berry-etal:07,siri-berry-etal:08,naude-cessac-etal:13}.  

In view of the present paper, the interest of chaos is that a linear response theory can be formally obtained in this case, despite, and actually, thanks to, the fact that dynamics is chaotic. This apparent paradox is discussed below.
We will consider a situation where chaos occurs via a  transition by quasi-periodicity  \cite{ruelle-takens:71}. When increasing $g$ out of the contractive regime the fixed point $\vVa$ destabilises by a Hopf bifurcation. As $g$ increases a second Hopf bifurcation generates a $2$ torus densely covered by trajectory. Then, frequency locking arises when crossing Arnold tongues \cite{arnold:83}. In the region where Arnold tongues overlap (edge of chaos) one can see succession of periodic and quasi-periodic windows until the appearance of a chaotic regime (strange attractor), in general by a period doubling cascade although other scenarios as possible (see \cite{mackay-tresser:86,gambaudo-tresser:88}, for a description of the possible scenarios). Numerical measurement of the Lyapunov spectrum exhibit one or more positive Lyapunov exponents in this regime \cite{cessac:95,siri-berry-etal:08}, corresponding to initial conditions sensitivity. The Lyapunov spectrum allows compute the dimension of the attractor via Kaplan-Yorke formula \cite{kaplan-yorke:79}. This dimension is not integer corresponding to a fractal attractor. Finally, from the Newhouse-Ruelle-Takens theorem \cite{newhouse-ruelle-etal:78} the attractor obtained at the end of this cascade of bifurcations is arbitrary close to an Axiom A  attractor. We come back to this last point below.

The dynamical regimes resulting from the increase in the gain parameter obviously depend on the form of the matrix $\cJ$. A typical situation where this transition to chaos occurs is when the synaptic weights $J_{ij}$ are random, independent, with 
Gaussian entries $\cN(0,\frac{J^2}{N})$. The parameter $J$ controls therefore the variance of the distribution. 
This case has been proposed by H. Sompolinsky and co-workers in a seminal paper dating back to 1988 \cite{sompolinsky-crisanti-etal:88}. 
 They considered the case $f(x)=\tanh(gx)$ for the continuous time model \eqref{eq:AWC}, so that the destabilisation of the stable fixed point $\vVa$ arises when $-1+g \Re(s_1)=0$ where $s_1$ is the (random) eigenvalue of $\cJ$ with the largest real part. 
Now, the asymptotic distribution of the spectrum ($N \to +\infty$) of this family of random matrices is known from a theorem due to Girko \cite{girko:84,girko:12}. The spectral density converges to a uniform distribution on the disk of center $0$ and radius $J$ in the complex plane. Thus, in the limit $N \to +\infty$ ("thermodynamic" limit), the fixed point destabilises for $gJ=1$. The same holds for the discrete time model \eqref{eq:AWC_Discrete}. It holds as well when breaking the symmetry $f(x)=-f(-x)$ although the bifurcations map is more complex \cite{cessac-doyon-etal:94}.

The spectral behaviour gives actually a qualitative way to explain why chaos is occurring in this case \cite{doyon-cessac-etal:93,cessac-doyon-etal:94}. When $N$ is finite, while increasing $g$, eigenvalues of the Jacobian matrix cross the boundary of dynamical stability by complex conjugate pairs, leading to the transition to chaos by quasi-periodicity. As $N$ increases the range of $g$ values where this cascade of transitions occurs shrinks to $0$. In the thermodynamic limit, infinitely many eigen-modes simultaneously destabilise when $gJ=1$, leading to infinite dimension chaos \cite{cessac:95}. Note that this scenario extends a priori to more complex forms of synaptic weights matrices, typically, having correlations, although this has not yet been really explored. 

The choice for the mean and variance scaling of the synaptic weights is  essential here: the variance scaling ensures that the sum of synaptic inputs have bounded variations as $N$ growths, ensuring a proper, non trivial, thermodynamic limit  $N \to +\infty$
; having a zero mean ensures that dynamics is fluctuations-driven (neurons are mostly in region II of figure \ref{fig:ExpCont}) leading to the so-called "balanced state" \cite{vanvreeswijk-sompolinsky:98}.

Random neural networks with random independent entries have attracted a lot of activity since the work  \cite{sompolinsky-crisanti-etal:88}. Especially, H. Sompolinsky and A. Zippelius \cite{sompolinsky-zippelius:81,sompolinsky-zippelius:82} developed an efficient dynamic mean-field approach for spin-glasses, which was later used to analyze the model \eqref{eq:AWC}. Although this theory is one of the most beautiful I know in the field of theoretical neuroscience (see \cite{schuecker-goedeke-etal:16} for a recent review) it requires a thermodynamic limit and deals with the average behaviour (weak convergence) of networks in this limit (although almost-sure convergence results now exist 
\cite{faugeras-maclaurin:15}) rendering difficult the interpretation of the dynamic mean-field equations and their solutions. Additionally, it relies heavily on the independence assumption of $J_{ij}$ (although large deviations techniques now allow to access correlated weights \cite{faugeras-maclaurin:15}).

In contrast, the study proposed here deals with a given network with a finite size. Although the numerical examples presented in this paper were generated by a random model with independent $\cN(0,\frac{J^2}{N})$ synaptic weights, the derivation of the linear response does not rely on this assumption. We do not average on the distribution of synaptic weigths, and we don't take the thermodynamic limit. As we see below, a nice resonances structure is produced by finite size models that is washed out by the mean-field approach (see \cite{muscinelli-gerstner-etal:18} for a recent result on resonances in mean-field theory of chaotic network, including plasticity. The resonance structure is quite different from what is obtained in finite networks). Beyond neuronal networks, this resonance structure has important consequences in the field of dynamical systems and statistical physics of chaotic systems, as discussed below.

\begin{figure}
\includegraphics[width=8cm, height=6cm]{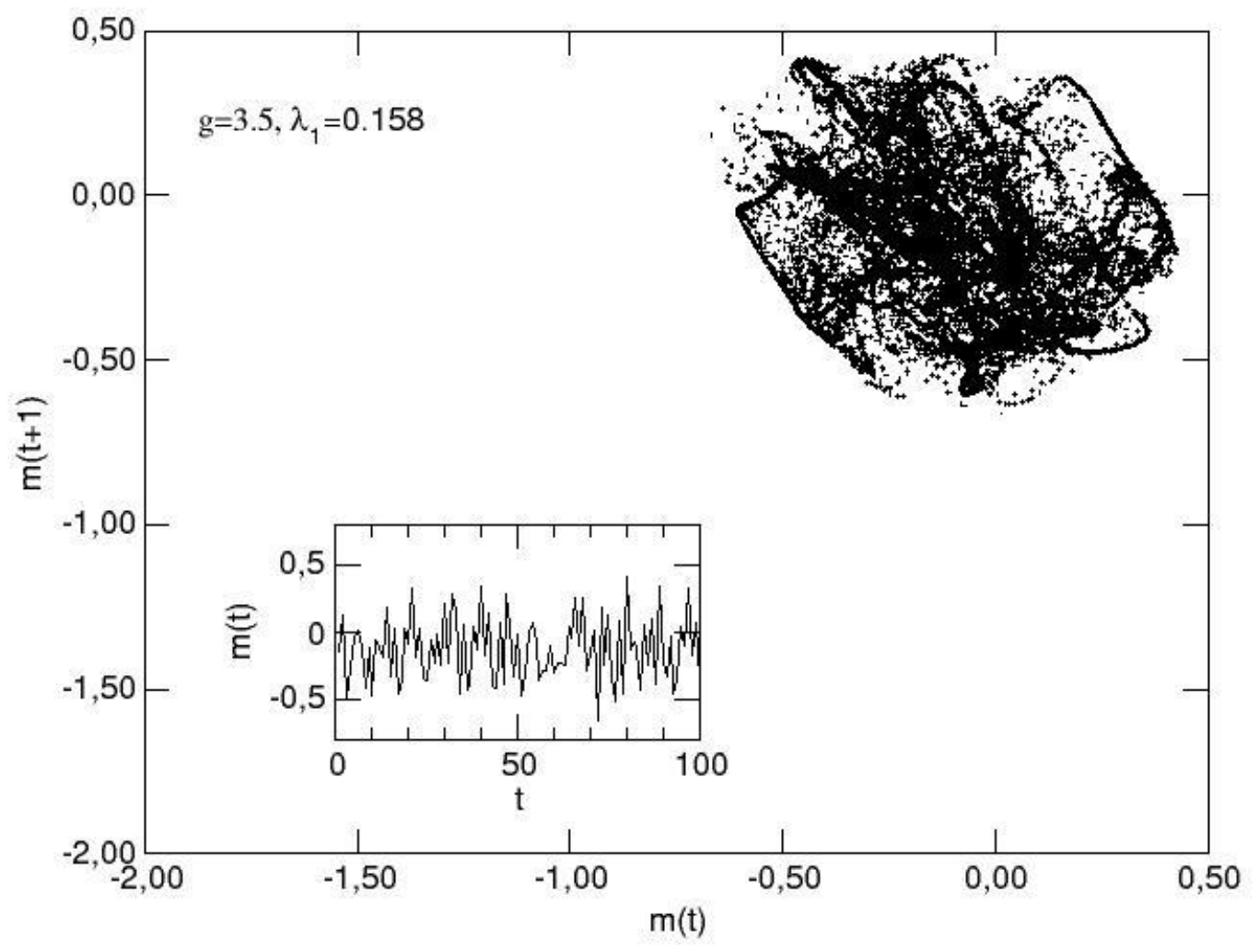}

\includegraphics[width=8cm, height=6cm]{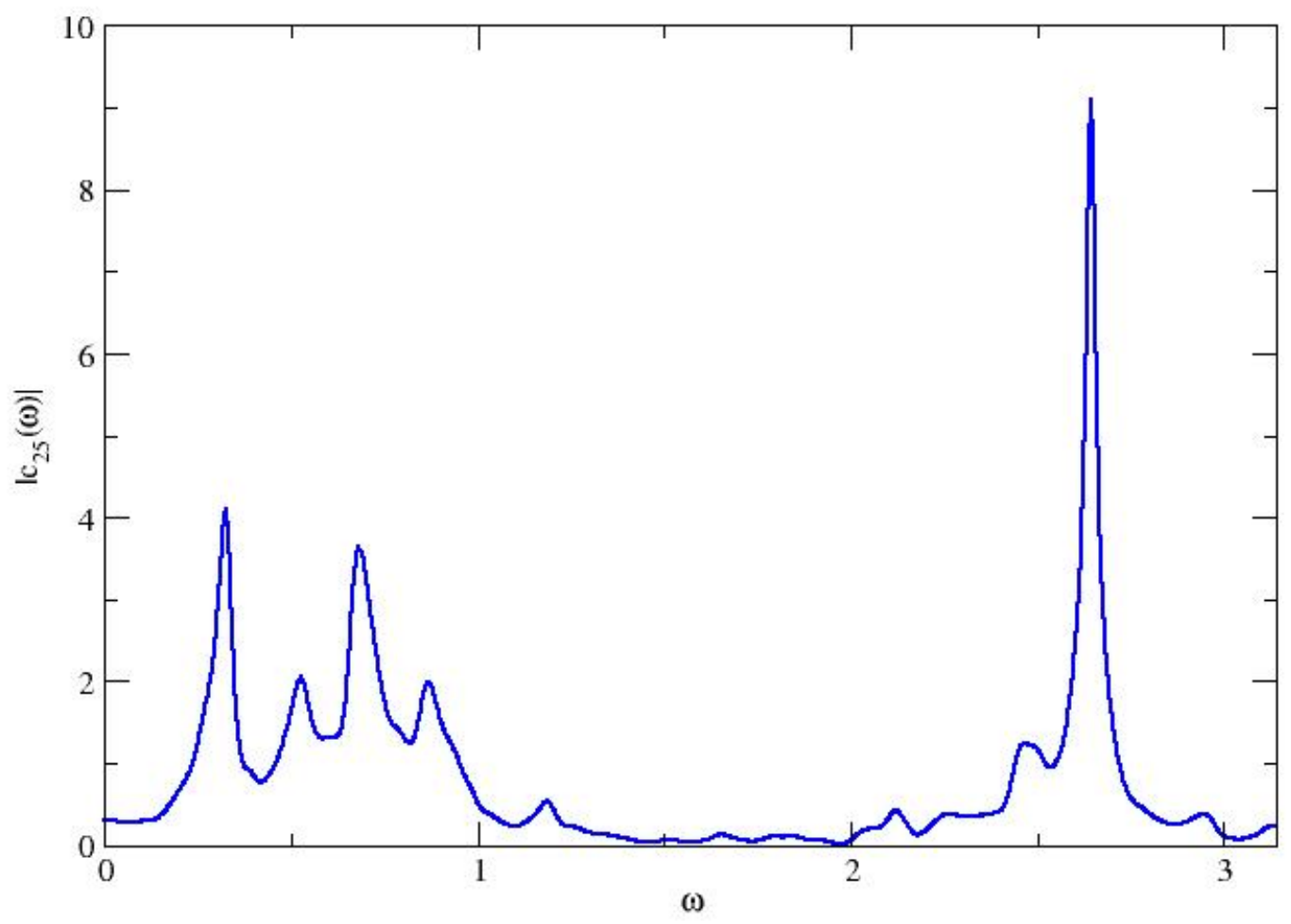}
\caption{\label{fig:ChaosFFT} Top. Representation of a strange attractor generated by \eqref{eq:AWC_Discrete}. This is obtained by plotting $m(t+1)$ versus $m(t)$ where $m(t)=\frac{1}{N} \, \sum_{j=1}^N V_j(t)$. Bottom. Power spectrum of $m(t)$, $\omega$ is the frequency (between $[0,\pi]$ as time is discrete).}
\end{figure}

In the next steps we are going to consider the discrete time version \eqref{eq:AWC_Discrete}. The main reason for this is that it affords an easy computation of the linear response in the chaotic regime, with a simple interpretation. An example of chaotic attractor obtained with \eqref{eq:AWC_Discrete} is given in Fig. \ref{fig:ChaosFFT} (top). In this regime the power spectrum of voltages (bottom) is continuous, but not flat, in contrast to white noise. There are peaks in the spectrum, corresponding to resonances in dynamics, called Ruelle-Pollicott resonances \cite{pollicott:85,ruelle:87,parry-pollicott:90,gaspard:05}, as discussed below. These resonances are independent of the neuron. They are reminiscent of fig. \ref{fig:Resonances_FP} although dynamics here is quite more complex. The peaks are related to the succession of bifurcations (two Hopf bifurcations and frequency locking) leading to chaos. 
  
\subsection{Linear response in the chaotic regime}\label{sec:LinRepChaos}

\subsubsection{$\epsilon$-expansion.}\label{sec:epsilon}

We now consider two versions of the dynamical system \eqref{eq:AWC_Discrete_vect}. The spontaneous dynamics version:
$$\vV(t+1) = \vG\pare{\vV(t)},$$
and the perturbed version:
\begin{equation}\label{eq:PerturbedMap}
\vV'(t+1) = \vG\pare{\vV'(t)} + \epsilon \vS(t).
\end{equation}
%
We assume that the stimulus is switched on at time $t_0$ so that $\vV(t_0)=\vV'(t_0)$.

We define
$\vdel(t) = \vV'(t) - \vV(t)$
the difference between the trajectories of the two systems. We have thus $\vdel(t_0+1)=\vV'(t_0+1) - \vV(t_0+1) = \epsilon \vS(t_0)$.
At time $t_0+2$:
$$
\vdel(t_0+2) 
%
=
\vG\pare{\vV(t_0+1)+\epsilon \vS(t_0)} + \epsilon \vS(t_0+1) -\vG(\vV(t_0+1))
$$
We now make a Taylor expansion of $\vG\pare{\vV(t_0+1)+\epsilon \vS(t_0)}$ in powers of $\epsilon$:
$$
\vdel(t_0+2) = \epsilon \bra{DG_{\vV(t_0+1)}.\vS(t_0)  +  \vS(t_0+1)} + \epsilon^2 \veta(t_0+1),
$$
where $\veta(t_0+1)$ contains terms of degree higher than $\epsilon$. Note that we don't assume that $\epsilon^2 \veta(t_0+1)$ is negligible so the equation is exact.

Iterating this procedure for larger times we obtain:
\begin{equation}\label{eq:Expansion_dV}
\vdel(t) = \epsilon \sum_{\tau=t_0}^{t-1} DG^{t-\tau-1}_{\vV(\tau+1)}.\vS(\tau)  + \epsilon^2 \vR(t)
\end{equation}
where, again, we do not assume that $\epsilon^2 \vR(t)$
is small and negligible.

\begin{figure}
\center
\includegraphics[width=8cm, height=7cm]{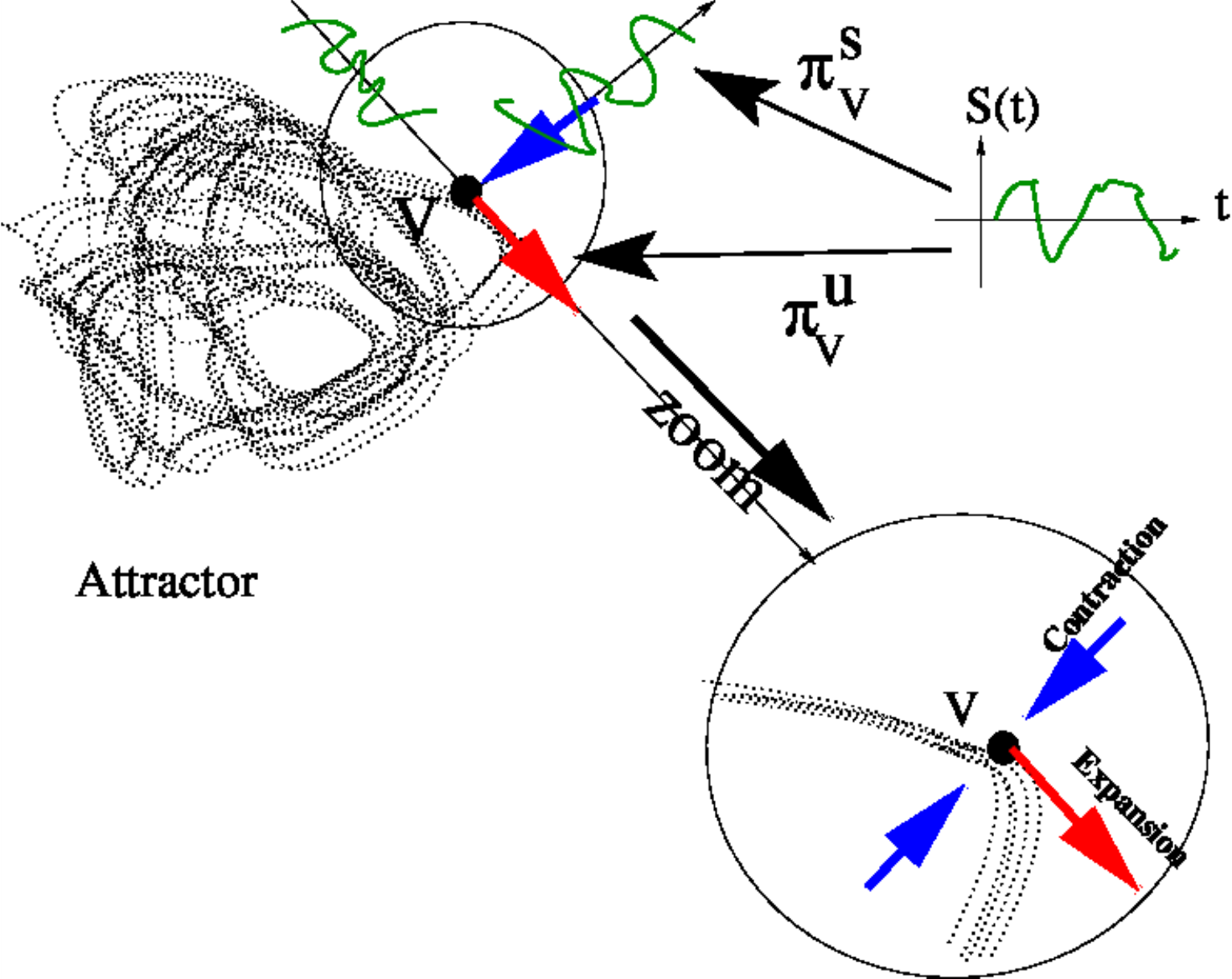}
\caption{\label{fig:ExpansionContractionAtt} The trajectory of \eqref{eq:AWC_Discrete_vect} following the attractor of the spontaneous dynamics is perturbed by the time dependent stimulus $S(t)$ (green). The stimulation projects on the local stable manifold (projection $\pi^s_{\vV}$) and is contracted; it also projects on the local unstable manifold (projection $\pi^u_{\vV}$) and is expanded. The directions of projections depend on the point $\vV$ on the attractor.}
\end{figure}

This formula, which generalises \eqref{eq:Chi_Stable_Disc}, looks a bit useless. Indeed, a linear response theory would neglect the term $\epsilon^2 \vR(t)$. But, in contrast to the contractive regime where the eigenvalues of $DG^{t-\tau-1}_{\vVa}$ had a modulus $< 1$, ensuring the convergence of the series, here,  the higher orders cannot be neglected, precisely because the system is chaotic. An initial perturbation, as tiny as it is, is locally amplified by dynamics. More precisely, 
dynamics is expansive in directions tangent to the attractor (positive Lyapunov exponents) and contractive in directions transverse to the attractor (negative Lyapunov exponents). This is illustrated in Fig. \ref{fig:ExpansionContractionAtt}. The sum of all Lyapunov exponents is negative expressing that volume is contracted in the phase space.
Asymptotic spontaneous dynamics lives on the attractor, so that a small perturbation - in our case, the stimulus $\epsilon \vS(t)$ - has generically a component tangent to the 	attractor and a component transverse to the attractor. The transverse component is exponentially damped, the tangent component is exponentially amplified.
Thus, the net effect is an amplification of the stimulus effect, ruining any hope to neglect the "residual" term $\epsilon^2 \vR(t)$ in \eqref{eq:Expansion_dV}. Therefore, it seems impossible to obtain a linear response on long times unless taking ridiculously small perturbations. This is the essence of the Van Kampen objection \cite{kampen:71}.

\subsubsection{Expansive dynamics and ergodic average} \label{eq:Ergodic}

\begin{figure}
\center
\includegraphics[width=7cm, height=5cm]{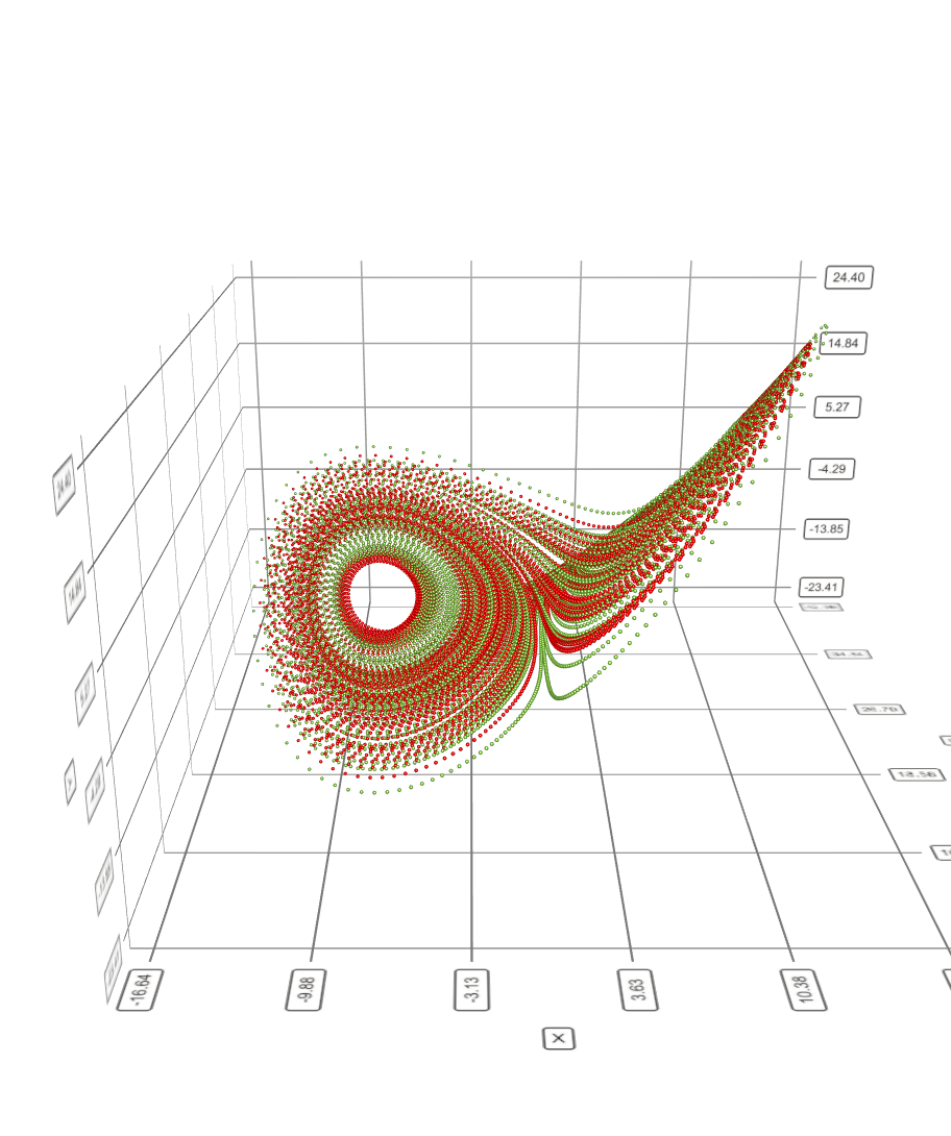}

\vspace{0.5 cm}

\includegraphics[width=7cm, height=5cm]{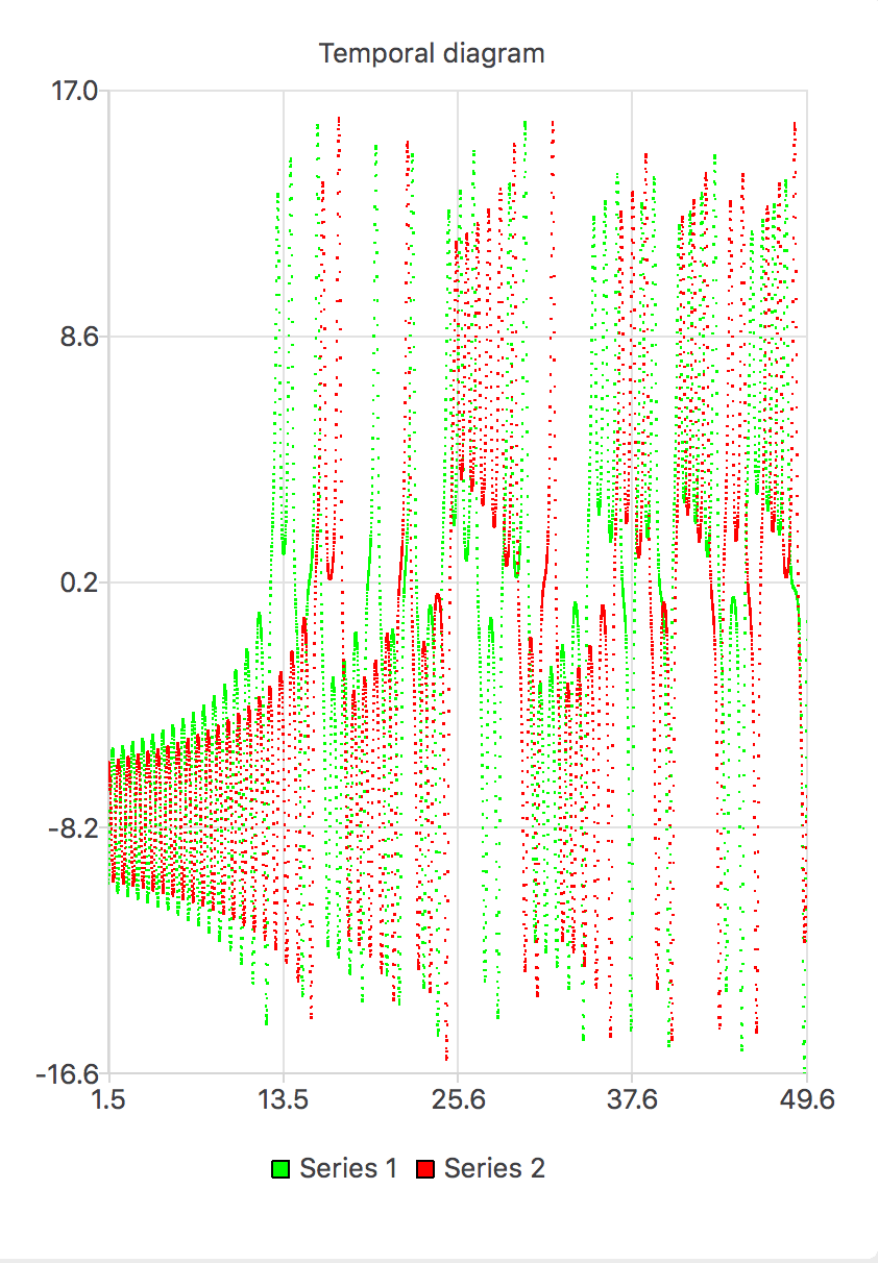}
\caption{\label{fig:LorenzSRB} Illustration of ergodicity on Lorenz attractor (top). We plot a trajectory (red) and a small perturbation of it (green). The green and red attractor look similar. Bottom. Temporal evolution of the $2$ trajectories. Although initial condition sensitivity initially separates the two trajectories, they mix (without crossing in the 3 dimensional space) after a certain time. }
\end{figure}

To make one step further, let us now consider in more detail the effect of a small perturbation in a celebrated example, the Lorenz attractor \cite{lorenz:63}. In fig. \ref{fig:LorenzSRB} we have represented a trajectory (in red) and a small perturbation of the red trajectory (in green). One clearly sees the initial condition sensitivity: the two trajectories initially diverge exponentially fast. However, because the phase space is compact, non linear folding takes place and the two trajectories get closer (they can get arbitrary close from Poincaré's recurrence theorem), without crossing though (from Cauchy's theorem on uniqueness of solutions). After a sufficiently long time it becomes impossible to distinguish the $2$ trajectories; the "green" attractor looks very much like the "red" one.

This example, illustrated here with the famous continuous time Lorenz model, illustrates a deep property that we are going to use now, ergodicity. This property holds, \textit{mutatis mutandis} for our discrete time chaotic system. 
 Consider, in our model, an initial condition $\vV$ and its trajectory $\vG^t(\vV)$, $t \geq 0$; consider a function $\Phi: \setR^N \to \setR^K$ (observable). Then, the time average of $\Phi$ on the trajectory, defined by the limit $\lim_{T \to \infty} \frac{1}{T} \sum_{t=1}^T \Phi(\vG^t(\vV))$, exists for typical initial conditions. We define now what we mean by "typical".

There exist a probability measure $\mu$ with support on the strange attractor $\Omega$, such that:
\begin{equation}
\lim_{T \to \infty} \frac{1}{T} \sum_{t=1}^T \Phi\left[\vG^t(\vV)\right] \stackrel{\mu \ a.s.}{=} \int_\Omega \Phi(\vV) 
\mu(d\vV)
\end{equation}
for $\mu$-almost every initial condition $\vV$. The time average of $\Phi$ along orbits is equal to the average of $\Phi$ with respect to $\mu$ (average on the attractor $\Omega$).

There exist a class of dynamical systems, called uniformly hyperbolic or axiom A (discussed in more detail below), for which the measure $\mu$ is obtained by the weak limit:
\begin{equation}\label{eq:SRB}
\mu \stackrel{w}{=} \lim_{t \to +\infty} \vG^{t} \mu_L, 
\end{equation}
where $\mu_L$ is the Lebesgue measure on the phase space $\cM$, and $\vG^{t} \mu_L$ is the image of $\mu_L$ by $\vG^{t}$. Such a measure is called the Sinai-Ruelle-Bowen measure \cite{sinai:72,bowen:75,ruelle:76,ruelle:78}.

For the SRB measure the following holds:
\begin{equation}\label{eq:SRB_Average}
\lim_{T \to \infty} \frac{1}{T} \sum_{t=1}^T \Phi\left[\vG^t(\vV)\right] \stackrel{\mu_L \ a.s.}{=} \int_\cM \Phi(\vV) 
\mu(d\vV) \equiv \brk{\Phi}_{eq}.
\end{equation}
This relation expresses that the time average of the trajectory of a "typical" initial condition, namely, selected with a uniform probability (Lebesgue measure) in $\cM$, or any probability having a density with respect to $\mu_L$, e.g. Gaussian, is equal to the average with respect to the SRB measure carried by the strange attractor. 

\subsubsection{Equilibrium versus non equilibrium}\label{sec:eq_non_eq}

This has very deep physical meaning, reflecting an intuitive notion, somewhat initiated by Boltzmann who invented the word "ergodic" \cite{gallavotti:14}. 
Starting from a "typical" microstate $\vV$, i.e. selected with a natural probability, e.g. uniform or Gaussian, the time average of an observable $\Phi$ along the dynamical evolution of $\vV$ is equal to the ensemble average of $\Phi$ with respect to the probability measure (macrostate) $\mu$. The SRB measure $\mu$ plays therefore the role of the Boltzmann-Gibbs distribution in statistical physics, but it is constructed from the dynamical evolution without need to invoke the maximum entropy principle. We refer to the unperturbed situation as an equilibrium situation. This is the reason why we used the subscript $eq$ in equation \eqref{eq:SRB_Average}. 

We want however to make here a small remark on the terminology. The theory used here, developed by Gallavotti, Cohen, Ruelle among others, was initially intended to characterise, from a dynamical system perspective, thermodynamic systems initially at equilibrium and perturbed by external forces, where the excess of energy is dissipated with a thermostat, ensuring a non equilibrium steady state. In this context, the equilibrium state has a density with the phase volume measure (the Liouville measure), while the non equilibrium state is characterised by a time dependent SRB measure which is not absolutely continuous. In contrast, our equilibrium state is a SRB measure, not absolutely continuous with respect to the volume element (it is only absolutely continuous along the unstable manifold). The theory developed by these authors extends to our case though \cite{ruelle:99}. 

Being ergodic SRB is stationary (time-translation invariant). In addition, the SRB measure is a Gibbs distribution whose "energy' is known and has the form:
\begin{equation}\label{eq:PotentielSRB}
H(\vV)=-\log \det \pi^u_{\vV} D\vG_{\vV};
\end{equation}
where $\pi^u_{\vV}$ is the local projection on the unstable manifold. The average energy with respect to the SRB measure, $\brk{H}_{eq}$, is the sum of positive Lyapunov exponents. The SRB measure obeys a maximum entropy principle and its entropy is the sum of positive Lyapunov exponents \cite{bowen-ruelle:75}. This is  Pesin formula \cite{pesin:77}.

An immediate consequence of ergodicity, somewhat appearing visually in Fig. \ref{fig:LorenzSRB}, is that the time average of $\Phi$ on a perturbed trajectory is equal to the time average of the unperturbed trajectory:
\begin{equation}\label{eq:EqualityAverages}
\lim_{T \to \infty} \frac{1}{T} \sum_{t=1}^T \Phi(\vG^t(\vV)) = \lim_{T \to \infty} \frac{1}{T} \sum_{t=1}^T \Phi(\vG^t(\vV + \vdel)).
\end{equation}
This essentially expresses that time average  smooths out the initial condition sensitivity and suggests to define a linear response theory via a proper averaging. Thinking of non equilibrium statistical physics this is exactly what we need. When considering particles in a fluid submitted to gradient of temperature, it is clear that molecular chaos and initial conditions sensitivity holds at the level of particles. But, at the level of a population, ensemble average, an order emerges, expressed by Fourier law, where the transport coefficient is expressed via correlations of flux computed at equilibrium from the Gibbs distribution. 

However, to define linear response in this context, one needs to extend the stationary situation exposed in this section, to a non stationary situation where the map defining the dynamics depends on time.

In this context, it is possible to define a time-dependent SRB measure by\cite{ruelle:99} (using our notations):
\begin{equation}\label{eq:TimeDependentSRB}
\mu_t= \lim_{n \to +\infty} \vG'_t  \dots  \vG'_{t-n} \, \mu_L,
\end{equation}
where $\vG'_t$ is the time-dependent map defined in \eqref{eq:PerturbedMap}. Equivalently, for an observable $\Phi$ the quantity:
\begin{equation}\label{eq:TimeDependentAverageSRB}
\brk{\Phi}_t = \int_\cM \Phi(\vV) \mu_t(d\vV)
\end{equation}
is the average value of the observable $\Phi$ at $t$, in the perturbed time-dependent evolution \eqref{eq:PerturbedMap}. 

\subsubsection{Linear response theory}\label{sec:RuelleLinRep}

D. Ruelle has developed a linear response theory for uniformly hyperbolic dynamical systems that we are going to use here \cite{ruelle:99}. That is, we  assume for the moment that our chaotic attractor is uniformly hyperbolic. We come back to this point in section \ref{sec:UnifHyp}.

We note $\delta_t \moy{\Phi}=\brk{\Phi}_t - \brk{\Phi}_{eq}$ the difference between the average of $\Phi$, at time $t$, for the perturbed time-dependent system and the average of $\Phi$ in the unperturbed system. This is the response of the perturbed system at time $t$, for the observable $\Phi$ and the stimulus $\vS$. 
Ruelle formula reads, in our case\cite{cessac-sepulchre:04,cessac-sepulchre:06,cessac-sepulchre:07}:
\begin{equation}\label{eq:RuelleRepLinGen}
\delta_t\moy{\Phi}= \epsilon \, \sum_{\tau=-\infty}^{t-1} \int \mu(d\vV)  \,
D\vG^{t-\tau-1}_{\vV}\vS(\tau). \vec{\nabla}_{\vV(t-\tau-1)}\Phi \, + O(\epsilon^2)
\end{equation}
(actually, Ruelle formula extends to cases where the stimulus depends on the state $\vV$).

Let us comment this formula in the simple case where $\Phi(\vV)=\vV$ so that $\vec{\nabla} \Phi=\vec{1}$. Then, $\delta_t\moy{\Phi}=\delta_t\moy{\vV}$ is the difference between the average voltage at time $t$ for the perturbed system and the unperturbed average. This gives;
\begin{equation}\label{eq:DefChi}
\delta_t\moy{\vV}= \epsilon \, \sum_{\tau=-\infty}^{t-1} \chi(t-\tau-1).\vS(\tau)  \, + O(\epsilon^2)
\end{equation}
where :
\begin{equation}\label{eq:chi_erg}
\chi(t-\tau-1)=\int \mu(d\vV)  \,
D\vG^{t-\tau-1}_{\vV}=\brk{D\vG^{t-\tau-1}}_{eq}
\end{equation}
is the linear response matrix. It is defined as 
the average of the Jacobian matrix $D\vG^{t-\tau-1}$ with respect to the equilibrium SRB state $\mu$. 
Note that \eqref{eq:RuelleRepLinGen} is a discrete time convolution, so that we can rewrite it in the form:
\begin{equation}\label{eq:LinRepChaosConv}
\delta_t\moy{\vV}= \epsilon \, \bra{\chi \ast S}(t),
\end{equation}
similar to \eqref{eq:SolLinStablePermanentDiscrete}. Thus, eq. \eqref{eq:chi_erg} provides an explicit form for the linear response kernel $K=\epsilon \chi$ in the chaotic regime of the discrete time Amari-Wilson-Cowan model.

Let us now compare equation \eqref{eq:RuelleRepLinGen} to equation \eqref{eq:Expansion_dV} obtained by a naive Taylor expansion where we had no hope to neglect the residual term $\vR(t)$, which actually increases in time due to the positive Lyapunov exponents. In contrast, here the residual term $O(\epsilon^2)$ remains under control and tends to zero like $\epsilon^2$ when $\epsilon \to 0$. 
Why is it so ? 

This is sketched in Fig. \ref{fig:ExpansionContractionAtt}. The stimulus perturbation locally projects on stable and unstable directions. In the stable direction, dynamics is contracting so perturbation is damped exponentially fast. In the unstable direction dynamics is expanding leading to amplification of a perturbation at the level of trajectories. However, considering averaging, as done in \eqref{eq:chi_erg}, the situation is different. It results indeed that the projection of the linear response on the unstable foliation is a correlation function between the observable $\Phi$ and a current $j_X=-\brk{div^u \vX}_{eq}$ where $div^u$ is the divergence computed along the attractor and $\vX$ a smooth perturbation. More precisely, one can define a local Riemmanian metric $\cG$ on the attractor so that the current reads \cite{ruelle:98}:
$$
j_X=-\frac{1}{\sqrt{\det \cG}} \sum_{i} \frac{\partial \pare{ \sqrt{\det \cG} \, X^i}}{\partial x^i},
$$
where $X^i$ are the coordinates of the projection of $\vX$ and $x^i$ the local coordinates on the attractor (expressed e.g. in the Lyapunov basis). 
In this context, Green-Kubo relation and Onsager coefficients can be computed from the entropy production \eqref{eq:EntropyProd}, not only at the lowest order, but also to higher orders \cite{gallavotti:96,ruelle:98,baiesi-maes:13,basu-maes:15}.\\
 
Correlation functions decay exponentially fast in chaotic (uniformly hyperbolic) systems (exponential mixing) ensuring the convergence of the series \eqref{eq:RuelleRepLinGen}. The decay rates are controlled by the eigenvalues of an evolution operator, acting on probabilities measures, the Ruelle-Perron-Frobenius operator, whose eigenvalues are the  Ruelle-Pollicott resonances. The projection of these complex poles on the real axis gives the peaks appearing in the power spectrum Fig. \ref{fig:ChaosFFT}. 

Thus, the cumulated effect of the stimulus along a trajectory, obtained via time averaging, does not diverge. It converges, on the stable foliation, because of volume contraction, and on the unstable direction because of exponential mixing.   

To sum up, the main difference between \eqref{eq:RuelleRepLinGen} and \eqref{eq:Chi_Stable_Disc} 
is averaging with respect to the equilibrium measure. This makes physical sense. As pointed above, there is no hope to characterise the response of a fluid to a temperature gradient at the level of a particles trajectory, but it is possible at the level of densities.

The link with non equilibrium statistical physics can be formally pursued further as discussed in the conclusion of this section.

\subsection{Linear response in the neural network model} \label{sec:LinRepFiringRate}

\subsubsection{Explicit form of the susceptibility}\label{sec:ExplicitForm}

In the discrete time model \eqref{eq:AWC_Discrete} the Jacobian matrix $D\vG_{\vV}= \cJ.\cD(\vV)$, where $\cD$ is defined in eq. \eqref{eq:diagD}. It is then easy to compute $\chi$ whose entry $\chi_{ij}$ reads:
\begin{equation}\label{eq:Chi_Discrete_Chaos}
\chi_{ij}(\sigma) = 
\sum_{\gamma_{ij}(\sigma)}
        \prod_{l=1}^{\sigma}J_{k_l k_{l-1}}
\brk{\prod_{l=1}^{\sigma}f'(V_{k_{l-1}}(l-1))}_{eq},
\end{equation}
where the sum holds on all synaptic paths connecting neuron $j$ to neuron $i$ in $\sigma$ steps, with $k_0=j$ and $k_\sigma=i$.

It is interesting to compare this equation to eq. \eqref{eq:chi_Contract} obtained in the contractive regime, where the attractor of dynamics was a fixed point. 
The straightforward difference is that we have now to average over the SRB measure the product of $f'(V_{k_{l-1}}(l-1))$. Actually, one obtains \eqref{eq:chi_Contract} by replacing the SRB measure by the Dirac measure on the attracting fixed point $\vVa$. This formula could be extended as well in the presence of noise.

Let us now interpret the meaning of the product $\brk{\prod_{l=1}^{\sigma}f'(V_{k_{l-1}}(l-1))}_{eq}$
weighting each path $\connect{j}{i}{\sigma}$. As we have seen in section \ref{sec:Contractive} the response of the post synaptic neuron $k_1$ to a small variation of the pre synaptic voltage $V_j$ which is proportional to $J_{k_1j} f'(V_j)$. Now, the main differences with the contractive regime are: (i) $V_j$ evolves in time; (ii) the gain $g>1$ can be quite high. This second aspect is essential because, near the inflexion point of the sigmoid, $f$ is expansive. It amplifies a small perturbation; in contrast it is contractive in the saturated parts (see Fig. \ref{fig:ExpCont}). This is actually precisely this interplay between expansion and contraction which is essential to render dynamics chaotic for sufficiently large $g$ (combined with the asymmetry of synaptic weights $J_{ij}$, as the model \eqref{eq:AWC_Discrete} with symmetric synapses has a Lyapunov function). 

\begin{figure}
\center
\includegraphics[width=6cm, height=5cm]{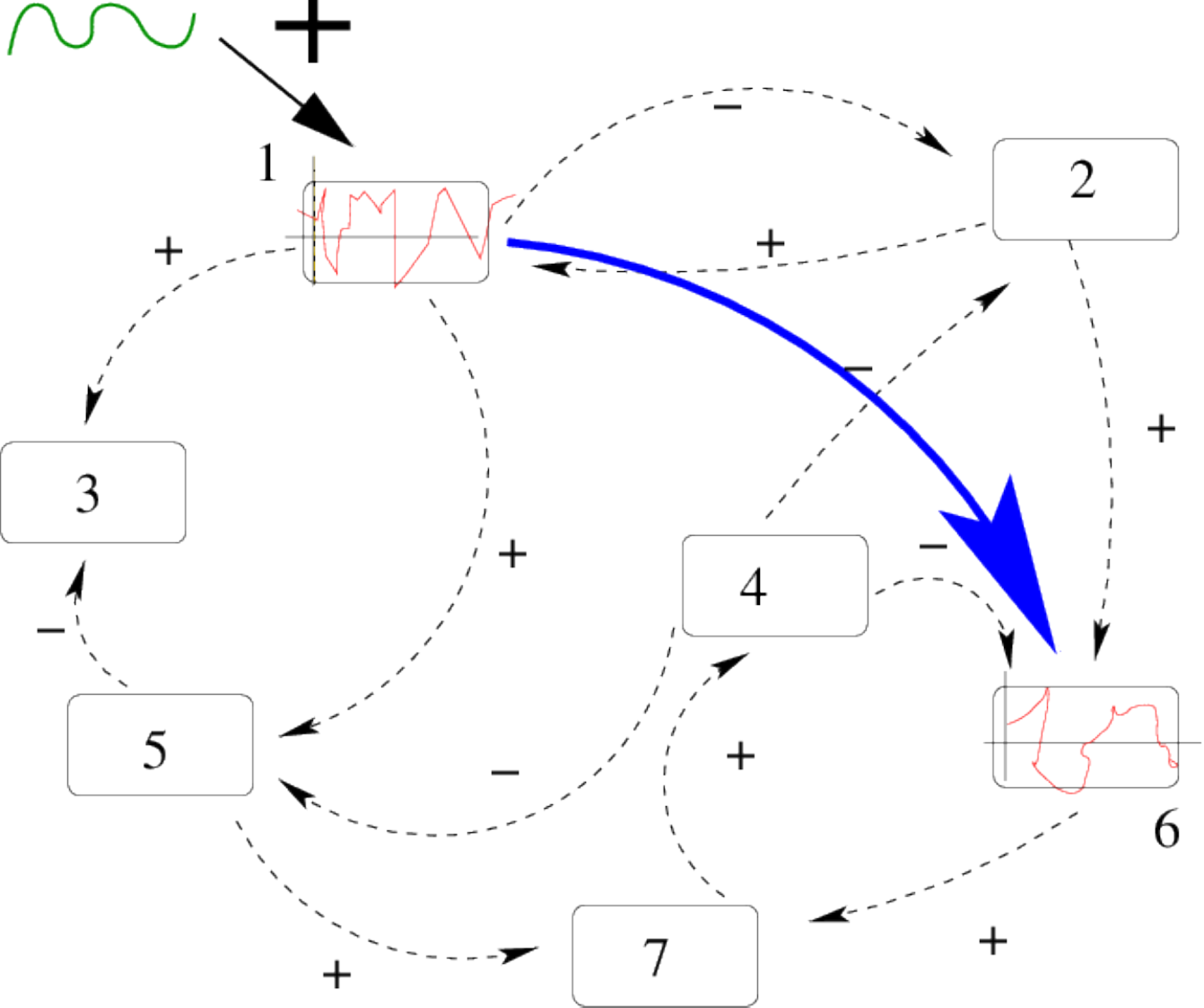}
\caption{Propagation of a stimulation throught the network in the chaotic case. In contrast to fig. \ref{fig:Circuits} the dynamics of the stimulated neuron ($1$) is now evolving chaotically (red trajectory) and the stimulus (green trace) is superimposed upon it. 
This affects the dynamics of neuron $6$ (blue arrow) but one has to disentangle the effect of the stimulus from the spontaneous activity in this neuron dynamics (red trace). 
   \label{fig:CircuitsChaos}}
\end{figure}

For clarity, let us consider, as in section \ref{sec:Contractive} the case when the signal is
injected only at neuron $j$ and let us study the response of neuron $i$. 
When the signal is injected at $j$, at time $t-\tau-1$, its propagation to $i$ via the network pathways is weighted by the products of terms $J_{k_lk_{l-1}} f'(V_{k_{l-1}}(l-1))$.
The contributions of all these paths sum up, with positive or negative weight (depending on the product of $J_{k_lk_{l-1}}$ along the path), with a small or large amplitude. The convolution form \eqref{eq:LinRepChaosConv} expresses that the response of neuron $i$ at time $t$ integrates the past influence of the stimulus injected at $j$, propagating via many paths with different lengths and summing up at $i$.

Now, the bracket $\brk{}_{eq}$ expresses 
that this influences is given by the ergodic average of  the products $\prod_{l=1}^{\sigma}f'(V_{k_{l-1}}(l-1))$. Remarkably, the averaging is done with respect to the equilibrium measure. We have here a strong analogy with non equilibrium statistical physics where transport coefficients are computed with respect to correlations functions of flux computed \textit{at equilibrium}, as exposed in section \ref{sec:StatPhys}. 

\subsubsection{Numerics}\label{sec:NumericsLinRepChaos}
The main problem with eq. \eqref{eq:Chi_Discrete_Chaos} is that it is numerically intractable. However, its Fourier transform is computable as we now explain. The (discrete time) Fourier transform of $\chi_{ij}(t)$ is:
$$
\hat{\chi}_{ij}(\omega)=\sum_{t=-\infty}^{+\infty} e^{i \, \omega \, t} \chi_{ij}(t).
$$ 

Consider now two perturbations, $\epsilon  \cos(\omega t) \ve_j$ and $-\epsilon  \sin(\omega t) \ve_j$ where $\ve_j$, is the canonical basis vector in direction $j$ and denote $\vV^{(1)},\vV^{(2)}$ the corresponding perturbed dynamics. Then, one can show that: \cite{cessac-sepulchre:04}
\begin{equation}\label{eq:Chi_ij_Numerical}
\hat{\chi}_{ij}(\omega)= \lim_{T \to + \infty} \frac{1}{T \, \epsilon} \sum_{t=0}^{T-1} e^{i \, \omega \, (t-1)} \bra{V^{(1)}_i(t) + i \, V^{(2)}_i(t) }.
\end{equation}
This allows the numerical computation of the complex susceptibility by time-averaging the trajectories of the two perturbed dynamics. Other numerical methods have been proposed in \cite{abramov-majda:08,baiesi-maes:13} that could be more efficient.

\subsubsection{Resonances} \label{sec:ResonancesChaos}

The computation of $\hat{\chi}$ allows to exhibit resonances, in a similar way as in section \ref{sec:Contractive}, but with a very different structure. First, these resonances are not given in terms of eigenvalues of the Jacobian matrix $D\vG_{\vV}$ because, in contrast to  section \ref{sec:Contractive}, $D\vG_{\vV}$ is now evolving dynamically along the strange attractor, as well as its eigenvalues. Rather than eigenvalues, Lyapunov exponents
express the average expansion/contraction rates, but I don't know about any result relating the Lyapunov spectrum to resonances. 
Here, resonances are obtained numerically using the form \eqref{eq:Chi_ij_Numerical}. 
We now interpret these resonances, first, in the context of dynamical systems theory and statistical physics, and then in the context of neuronal networks. \\

A classical wisdom coming from fluctuation-dissipation theorem in statistical physics is that the complex susceptibility is the Fourier transform of the corresponding correlation function, so that the resonances are peaks in the power spectrum. However, the
implicit assumption underlying this result is that the equilibrium distribution has a density with respect to the Lebesgue (or Liouville) measure.
 
The situation is more complex in the case of strange attractors because the SRB measure is absolutely continuous only along the unstable manifold (parallel to the attractor) and is singular (fractal) transverse to the attractor. As a consequence, Ruelle's theory asserts that the linear response operator is the sum of two contributions. There is a regular term, corresponding to the response to perturbations “parallel” to the attractor (locally projected along the unstable manifold). This term is a correlation function and, consequently obeys the Fluctuation-dissipation theorem. The poles of its Fourier transform are the Ruelle-Pollicott resonances. They give the rate of mixing of the chaotic system, or, equivalently, the relaxation rate to equilibrium for a perturbation “on” the attractor. These poles are independent of the observable. Thus, in our case, they are independent on the pre-synaptic, post-synaptic neuron pair. These resonances are observed in Fig. \ref{fig:ChaosFFT}.

\begin{figure}
\center
\includegraphics[width=8cm, height=6cm]{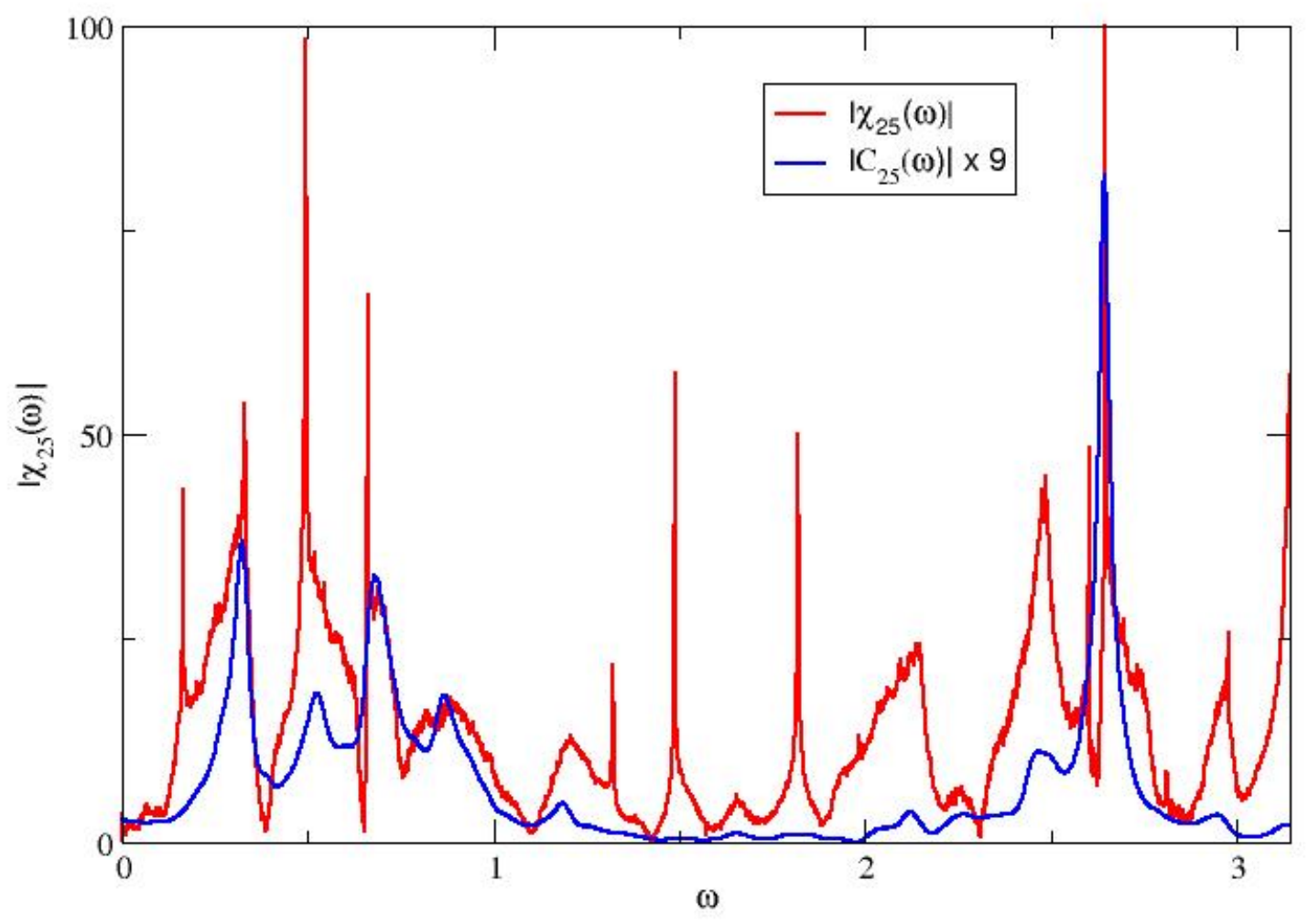}
\caption{\label{fig:Resonances} Resonances in the response of a neuron (here $2$) when exciting another neuron (here neuron $5$) with an harmonic stimulus. In red we have plotted the modulus of $\hat\chi_{25}$ (eq.  \eqref{eq:Chi_Discrete_Chaos}) as a function of the real frequency $\omega$. In blue is plotted the power spectrum of neuron $j$
(this is the same as in fig. \ref{fig:ChaosFFT}). The resonances which are not in the power spectrum are the stable resonances predicted by D. Ruelle \cite{}.}
\end{figure}

There is a second term in the linear response operator, corresponding to the response to perturbations locally projected along stable manifolds, namely transverse to the attractor. Therefore, this term exists only in the dissipative case. It does not obey fluctuation-dissipation theorem and its resonances have a different structure. These exotic resonances, theoretically predicted by Ruelle \cite{ruelle:99}, were, to my best knowledge exhibited for the first time by J.A. Sepulchre and myself in \cite{cessac-sepulchre:04}. That was in the model \eqref{eq:AWC_Discrete}. An example is shown in fig. \ref{fig:Resonances} where, in blue is plotted the power spectrum (with peaks corresponding to Ruelle-Pollicott resonances) and in red are plotted the resonances in the complex susceptibility.

From the point of view of neuronal dynamics resonances have the following interpretation. In the chaotic regime neurons have firing rates evolving in an unpredictable
and erratic way. Yet, there is an hidden dynamical structure revealed by the resonances in Fig. \ref{fig:Resonances}.  First, upon exciting neuron $j$ with a weak amplitude, harmonic signal, with resonant frequency $\omega$ superimposed upon the chaotic activity of neuron $j$, the effects of this stimulation will propagate through the network, and can actually be measured.
Indeed,  some neurons in the network will respond with a maximal amplitude (given by the modulus of $\hat{\chi}$). This effect can be exploited to transmit a signal with slow modulation through the network - despite chaos - using a resonance frequency as a carrier frequency.
This signal, scrambled in the chaotic dynamics, can nevertheless be recovered on specific nodes, using an appropriate average procedure. See \cite{cessac-sepulchre:06} for further details.

This leads to a define an effective connectivity (sketched by the blue arrow in Fig. \ref{fig:CircuitsChaos}) depending on the frequency.
It does not coincide with the synaptic graph. It does not coincide
either with a graph built on correlation functions precisely because the susceptibility
has a contribution which is not a correlation function.
This suggests that groups of neurons can synchronise upon excitation of the network with a suitable resonance frequency. These clusters of neurons would depend on the excitation frequency, supporting the idea that neurons can dynamically constitute far more clusters than expected from the study of the synaptic graph.  

\subsubsection{Uniform hyperbolicity} \label{sec:UnifHyp}

Ruelle's theory mathematically requires the dynamical system (at least the dynamics on the attractor) be uniformly hyperbolic. 
Uniform hyperbolicity means that for for every $\vV \in \Omega$ there is a splitting of the tangent space $T_{\vV} \Omega=E_s(\vV) \oplus E_u(\vV) $, and there are constants $C>0$ and $0 < \lambda <1$, such that, for every $n \in \setN$, one has $\|D\vG^n_{\vV}.\vu\| \leq C \, \lambda^n \|\vu\|, \quad \vu \in E_s(\vV)$
(exponential contraction on the stable space $E_s(\vV)$),
and $\|D\vG_{\vV}^{-n}.\vu\| \leq C \, \lambda^n \|\vu\|, \quad \vu \in E_u(\vV)$, (exponential expansion on the unstable space $E_s(\vV)$) \cite{katok-hasselblatt:98}. 
In a nutshell this means that there are no neutral eigendirections (eigenvalues with modulus $1$) for the derivative $D\vG$. 

Does the dynamical system  \eqref{eq:AWC_Discrete} obey
this condition ? This is a tricky question for which I  only have partial answers. First, remark that we are dealing here with \textit{random} attractors as each realisation of the $J_{ij}$ generates a different attractor. So the answer to this question would require a probabilistic treatment.  For a given realisation of $J_{ij}$s one can measure the Lyapunov spectrum  and check that there are positive Lyapunov exponents. But having positive Lyapunov exponents is not enough to guarantee uniform hyperbolicity because neutral points may exist, as e.g. in the Henon model. What is known from the Newhouse-Ruelle-Takens theorem \cite{newhouse-ruelle-etal:78} is that the attractor resulting from a transition to chaos by quasi-periodicity can be perturbed by a smooth perturbation to give rise to an axiom A (uniformly hyperbolic) attractor.  So our  attractor is, in this sense, "close" to uniform hyperbolicity, no more.
 From the shape of the Jacobian matrix and the form of the sigmoid, it is clear that there is a subset of the phase space where $D \vG_{\vV}$ has some eigenvalue with modulus $1$.
The question is whether the strange attractor $\Omega$ intersects this region. The answer depends on the matrix $\cJ$ and I don't know about any methods allowing to solve this question. 

Therefore, the derivation made for the Amari Wilson Cowan model in the chaotic regime are made "as if" this system was uniformely hyperbolic. This assumption is called "chaotic hypothesis". It has been proposed by Gallavotti and Cohen \cite{gallavotti-cohen:95,gallavotti-cohen:95b}.This conjecture agrees with the fact that many chaotic time evolutions behave as if they corresponded to uniformly hyperbolic dynamics.  

Note that there are known example of chaotic attractors with neutral points (like the Henon's attractor) where linear response is violated \cite{ruelle:05,jiang-ruelle:05,baladi:07,baladi-benedicks-etal:14}. The nice point is that this violation can be detected numerically \cite{cessac:07}. We have not observed such effects in the examples of model \eqref{eq:AWC_Discrete} that we have studied.

\subsection{Conclusion of section \ref{sec:LinRepFiring}}

In this section, we have been able to answer the questions raised in the introduction using a linear response theory developed for chaotic dynamical systems using a form of Gibbs distribution, the SRB measure. We have been able to compute the convolution kernel in terms of the parameters and function ruling the non linear neurons dynamics. We have characterised how a stimulation of weak amplitude, applied to a group of neurons, influences the whole network. The functional connectivity obtained this way, clumsy sketched by the blue arrow in Fig. \ref{fig:CircuitsChaos}, is quite different from the synaptic connectivity as it results from a complex interplay between the network structure, the non linear dynamics and the statistics of orbits characterised by the SRB measure.  


We want to address a few pending questions.\\

\paragraph{Information transport.} Can we relate the effective connectivity to some notion of information transport that could be relevant for neuronal networks ?
We stay here at a formal level and the following discussion would require further developments. A good candidate to quantify the effect of gently varying the voltage $V_j$ of neuron $j$ is the derivative $f'(V_j)$, or better, its log, $\log \, f'(V_j)$. Why the log ? Because the effect of $f'$ is multiplicative along trajectories and because it controls the exponential expansion/contraction rate along the orbits of the dynamics. Actually,  the average volume contraction rate in the discrete time Amari-Wilson-Cowan model is 
$\brk{\log \abs{\det D\vG}}_{eq} = \log \abs{\det \cJ}  + \sum_{j=1}^N \brk{\log f'(V_j)}_{eq}$. This average volume contraction, corresponds to an entropy production rate, as used in \cite{gallavotti:96,ruelle:99}. It has a static, synaptic network dependent term, and a dynamical term. From the analogy with 
\eqref{eq:EntropyProd}, one may define a formal current form neuron $j$ to neuron $i$ by:
\begin{equation}\label{eq:Currentj-i}
\vec{j}_{ij}=\frac{\partial \brk{\log f'(V_i)}_{eq}}{\partial V_j}.
\end{equation}

This involves the derivative of the SRB state which can be computed by \eqref{eq:RuelleRepLinGen}. This could be a formal way to define the blue arrow in Fig. \ref{fig:CircuitsChaos}. It remains however to check whether this formal current can be measured and how it relates to more classical indicators like information flow or Granger causality. \\

\paragraph{How large can be $\epsilon$ ?}
Linear response theory assumes that higher order terms are negligible. This assumption holds true for a range of $\epsilon$ values where the second order term is smaller than the first order. Thus, answering this question requires a priori compute the second order term. Although there is a general formulation of higher order terms \cite{ruelle:98}, their computation has not been done in the present context. In the presented work, the validity of the linear response was numerically done, checking that e.g. dividing $\epsilon$ by $2$ gives a response with an amplitude divided by $2$. The typical values for $\epsilon$ were of order $10^{-2}-10^{-3}$, thus small compared to the amplitude of the voltage. However, as pointed out in the introduction, linear response theory is only a first step toward characterising neural responses to stimuli. We further develop this point in the conclusion section.

\section{From spiking neurons dynamics to linear response}\label{sec:LinRepSpiking}

In this section, we address again the questions raised in the introduction, in a different context, for a spiking neuronal network. As most neurons communicate by spikes
there are many models of spiking neurons that we can roughly divide in two categories: (1) (Non) linear smooth dynamical systems, based on \eqref{eq:ChargeCons}, like Hodgkin-Huxley \cite{hodgkin-huxley:52}, FitzHugh-Nagumo \cite{fitzhugh:55,nagumo-arimoto-etal:62} or Morris-Lecar \cite{morris-lecar:81} where the spike is a continuous function of time, shaped by the ionic currents involved in \eqref{eq:ChargeCons}. These models are close to biophysics but remain quite hard to study at a theoretical level. For this reason researchers have developed a second category: (2) Non linear, non smooth dynamical systems, based on \eqref{eq:ChargeCons}, like the Integrate and Fire model \cite{lapicque:07} and its generalisations \cite{izhikevich:03,brette-gerstner:05,rudolph-destexhe:06} where the  spike is a discontinuous function of time. The voltage $V$ obeys equation \eqref{eq:ChargeCons} (sub-threshold dynamics), with simplified ionic current terms, until  $V$ reaches a threshold. Then it is instantaneously reset to a rest value and a spike is emitted. This class of model is simpler to study and simulate although the introduction of a discontinuity at the threshold has a price to pay (non smoothness). 
In the case of the linear response theory developed here, this requires to use rather different tools than the previous section.

Here, I present a summary of work done in collaboration with Rodrigo Cofr\'e where we developed a linear response theory in a conductance-based integrate and fire model \cite{cofre-cessac:13,cessac-cofre:13,cessac-cofre:19}. As in the previous section, we are able to derive a linear response formula where the response kernel is written in terms of spike correlations, with an explicit dependence in the networks parameters.
This work is in line with other approaches attempting to understand how the correlated spiking activity of a neuronal network reflects the neurons interactions as well as their collective response to stimuli. 
Modelers have proposed methods based on statistical physics (Maximum Entropy Models \cite{schneidman-berry-etal:06,shlens-field-etal:06,shlens-field-etal:09,ferrari-deny-etal:18}, stochastic processes (Markov chain, Hawkes processes \cite{reynaud-bouret-rivoirard-etal:13,reynaud-bouret-rivoirard-etal:14}) or phenomenological models (Linear-non linear, Generalized Linear Models \cite{chichilnisky:01,simoncelli-paninski-etal:04b,ostojic-brunel:11}) to achieve this purpose. 
Yet, there are rather few modelling studies trying to relate the collective dynamics of neurons with the spike statistics of the network, in spontaneous activity as well as in the presence of a stimulus.

\subsection{Model}

\subsubsection{Spike emission}

\begin{figure}
\center
\includegraphics[width=8cm, height=5cm]{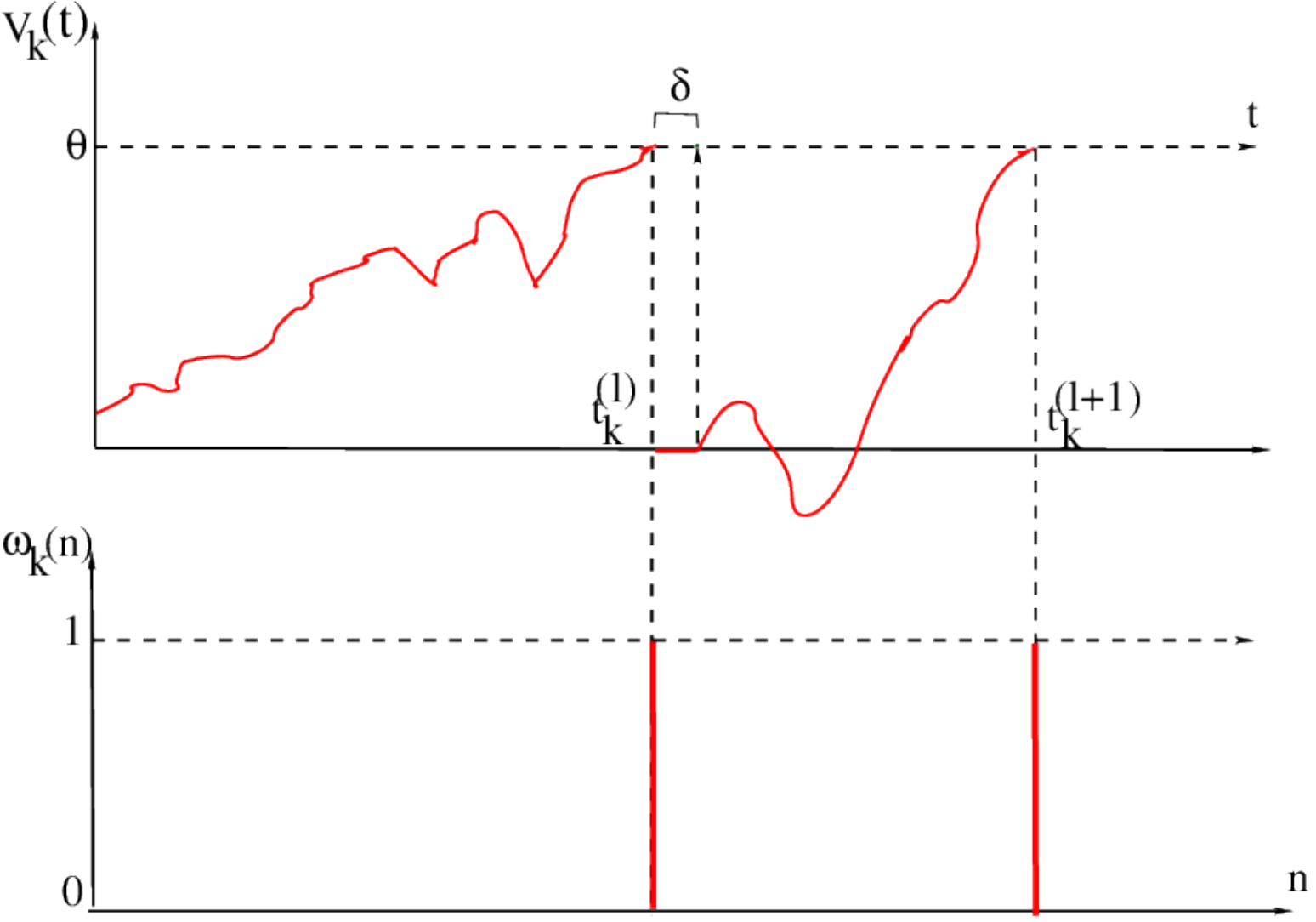}
\caption{\label{fig:IF} Integrate and Fire dynamics. When the voltage $V_k$ reaches the threshold $\theta$, at time $t_k^{(l)}$, it is reset to $0$ and a spike is recorded (Fire). Voltage stays at this value during a time interval $\delta$. Then evolution of $V_k$ starts again (Integrate). }
\end{figure}

As in the previous section we consider $N$ neurons with voltage $V_k$, $k=1 \dots N$. But dynamics is of a different nature because we focus here on spike statistics.
Spike emission is a complex process \cite{dayan-abbott:01}. In Integrate and Fire models this process is quite simplified. The principle is illustrated in fig. \ref{fig:IF}. 

We fix a voltage threshold $\theta$. Below threshold, $V_k$ follows and evolution based on eq. \eqref{eq:ChargeCons} (see eq. \eqref{eq:subthresholdVk} below). If the neuron $k$'s voltage reaches the threshold at time $t$ this neuron emits a spike. Then, its voltage is reset to a rest value (taken here to be $0$ without loss of generality) and the neuron stays at this value (quiescent) during a time interval $\delta>0$. 

We note $t_k^{(l)} $ the time of occurrence of the $l$-th spike emitted by neuron $k$. We define a  spiking variable $\omega_k(n) \in \Set{0,1}$ where $n$ is an integer. We set $\omega_k(n) = 1$ if neuron $k$ emits a spike in the time interval $[n \delta, (n+1) \delta[$ and $\omega_k(n) = 0$ otherwise. This reads:
\begin{equation}\label{eq:SpikeDiscretisation}
\omega_k(n)=
\left\{
\begin{array}{lll}
1,& \quad \mbox{if} \quad  \exists l, t_k^{(l)} \in [n \delta, (n+1) \delta[\\
0,& \quad \mbox{otherwise}.
\end{array}
\right.
\end{equation}
Spiking variables are therefore time-discrete events  with a time resolution $\delta$. 
We define the spike pattern of the network at time $n$ by the vector $\omega(n)=\vect{\omega_k(n)}_{k=1}^N$.
A spike block $\seq{\omega}{m}{n}$, $m < n$, is the matrix of spike patterns $\pare{\omega(m), \dots, \omega(n)}$. A spike train is a bi-infinite spike block $\seq{\omega}{-\infty}{+\infty}$. We note it $\omega$ for simplicity.

In this section we are going to consider a mixed dynamics with continuous time variables and discrete time variables. Especially, we will consider functions of the type $f(t,\omega)$ where $t$ is the continuous time variable and $\omega$ the discrete time spike train. In this notation, however, $f(t,\omega)$ signifies $f(t,\seq{\omega}{-\infty}{\ent{t}})$ where $\ent{t}$ is largest integer smaller or equal than $t$. This condition expresses causality: the function $f(t,\omega)$ depends on the spikes emitted \textit{before} time $t$.

\subsubsection{Sub-threshold dynamics}

The subthreshold dynamics of neuron $k$ is based on a model proposed by M. Rudolph and A. Destexhe in \cite{rudolph-destexhe:06}.
It starts from the charge conservation equation \eqref{eq:ChargeCons}, where the ionic current is a simple, passive leak term $-g_{L}(V_k-E_{L})$. The external current (stimulus) takes the form $\epsilon S_k(t)$ and the noise term reads $\sigma_B \xi_k(t)$ where $\xi_k(t)$ is a white noise. $\sigma_B$ controls the amplitude of the noise.

\begin{figure}
\center
\includegraphics[width=4cm,height=3cm]{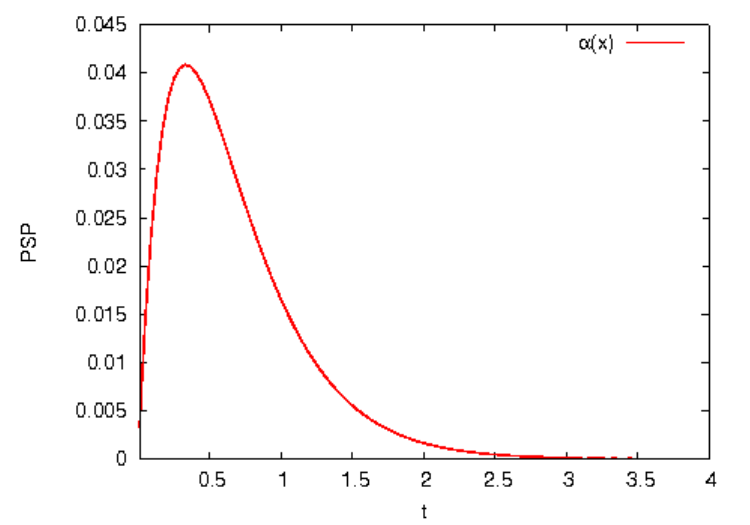}
\hspace{0.3cm}
\includegraphics[width=4cm,height=3cm]{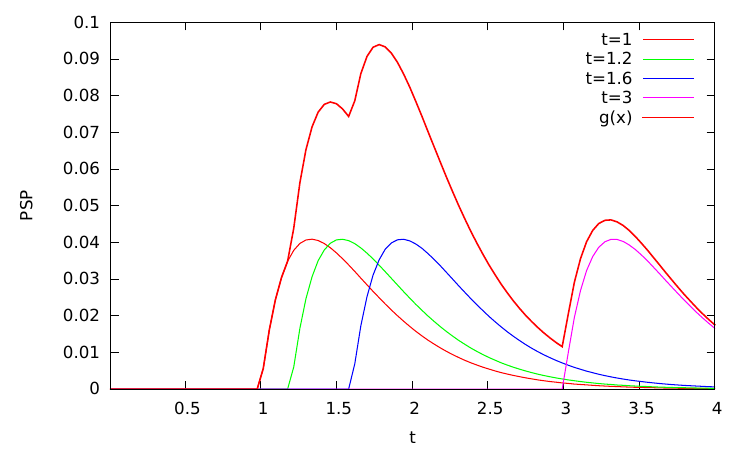}
\caption{Conductance of the model. Left. The function $\alpha$ of eq. \eqref{eq:alpha}.
Right. Spikes are emitted at times $1,3,12,16$, generating an increase in the conductance $g_{kj}$ controlled by the $\alpha$ function \eqref{eq:alpha}. The total conductance is the sum of the $\alpha$ profiles. 
\label{fig:alpha}}
\end{figure}

The synaptic current from pre-synaptic neuron $j$ to post-synaptic neuron $k$ reads $-g_{kj}(t,\omega)(V_k-E_{kj})$ where $E_{kj}$
is the reversal potential associated with the synapse $j \to k$.
In this model, the synaptic conductance $g_{kj}$ depends on time and on the previous spiking history of the pre-synaptic neuron $j$. Whenever neuron $j$ emits a spike (at time $t^{(n)}_j$) the conductance increases by an amount $G_{kj} \alpha (t - t^{(n)}_j)$, where $G_{kj}$ is the maximal conductance, and:
\begin{equation}\label{eq:alpha}
\alpha (t) = \frac{t}{\tau} e^{-\frac{ t}{\tau}} H(t),
\end{equation}
mimics the time profile in the synaptic increase upon emission of a pre-synaptic spike (Fig. \ref{fig:alpha}). Here, $H(t)$ is the Heaviside function. 

The total conductance is $g_{kj} (t) = G_{kj} \sum_{n \geq 0} \alpha (t - t^{(n)}_j)$. It depends therefore on the spike history, represented by the spike times $t^{(n)}_j$ preceding $t$. The set of all possible such times is uncountable. In order to have a dependence in a countable set of events we use the spike time discretisation \eqref{eq:SpikeDiscretisation} to obtain:
\begin{equation}\label{eq:gk}
g_{kj} (t,\omega) = G_{kj} \sum_{n \geq 0} \alpha (t - n \delta)\omega_j(n)
\end{equation}

Setting:
\begin{equation}\label{eq:Wkj}
W_{kj} = G_{kj} E_{kj},
\end{equation}
\begin{equation}\label{eq:alphakj}
\alpha_j(t,\omega) = \sum_{n \geq 0} \alpha (t - n \delta)\omega_j(n),
\end{equation}
\begin{equation}\label{eq:ik}
i_k(t,\omega) = g_{L} E_L + \sum_j W_{kj} \alpha_j(t,\omega) + \epsilon S_k(t) + \sigma_B \xi_k (t),
\end{equation}
we arrive at the final equation for the sub-threshold dynamics of neuron $k$:
\begin{equation}\label{eq:subthresholdVk}
C_k \, \frac{dV_k}{dt}+\gk{t,\omega} V_k=i_k(t,\omega), \quad \mbox{if} \quad V_k < \theta.
\end{equation}

Let us summarise. Spike time is discretised in time bins $n \delta$. Voltage, current and conductance time $t$ is continuous. When the voltage of neuron $k$, $V_k$, reaches the threshold, $V_k(t_k^{(l)} )=\theta$, with $t_k^{(l)}  \in [n \delta, (n+1) \delta[$ for some $n$ it is reset to $0$, and the $l$-th spike of neuron $k$ is recorded at discrete time $n$, $\omega_k(n)=1$. Voltage stays at $0$ until time $(n+1) \delta$ where it follows the sub-threshold evolution \eqref{eq:subthresholdVk} until the next time where $V_k$ reaches the threshold. Note that, in this modelling, $\delta$ can be quite small compared to the time scales of the dynamics.\\

\textbf{Remark.} Although eq. \eqref{eq:subthresholdVk} looks quite simple (it is a differential equation, linear in $V_k$, with time dependent coefficients) it hides a real complexity, the dependence in the history $\omega$. The coefficients $\gk{t,\omega},i_k(t,\omega)$ depends on the trajectory of $V_k$
which itself depends on the history of spikes anterior to $t$. This involves compatibility conditions between the trajectory and the spike train $\omega$ which constitutes a symbolic coding of trajectories. These aspects are further discussed in \cite{cessac:11b}.

\subsubsection{Solutions}

For a time $t$, a spike train $\omega$ and a neuron $k$ 
we note $\tko$ the last time anterior to $t$ where the neuron membrane potential was reset. 

It is easy to integrate the linear system \eqref{eq:subthresholdVk} from time $\tko$ to time $t$.
The corresponding flow is:
$$
\Gamma_k(t_1,t,\omega)=
\left\{
\begin{array}{lll}
e^{-\frac{1}{C_k}\int_{t_1}^{t}\gk{u,\omega} \, du}, \, &\mbox{if} \quad t \geq  t_1  \geq \tko;\\
0, &\mbox{otherwise} \,  .
\end{array}
\right.$$

We obtain:
\begin{equation}\label{eq:Vk_integ}
V_k(t,\omega)= \underbrace{\Vkspont{t,\omega} +  \Vkext{t,\omega}}_{\Vkdet{t,\omega}}+\Vknoise{t,\omega},
\end{equation}
where:
$$
\Vkspont{t,\omega}  =  \Vksyn{t,\omega} +\VkL{t,\omega},
$$
is the spontaneous contribution with:
$$
\Vksyn{t,\omega} \, = \, 
 \frac{1}{C_k} \sum_{j=1}^N   W_{kj} \, \int_{\tko}^{t} \Gamma_k(t_1,t,\omega) \, \alpha_j(t_1,\omega) \, dt_1,
$$
the synaptic interaction term, and,
$$
\VkL{t,\omega} \, = \,
\frac{E_L}{\tLk} \int_{\tko}^{t} \, \Gamma_k(t_1,t,\omega) \,  dt_1,
$$
where:
$$
\tLk \deq \frac{C_k}{g_L}.
$$
The second term in \eqref{eq:Vk_integ} corresponds to the contribution of the external stimulus:
\begin{equation}\label{eq:Vkext}
\Vkext{t,\omega} \, = \,
\epsilon  \, \frac{1}{C_k} \, \int_{\tko}^{t} S_k(t_1) \Gamma_k(t_1,t,\omega) \, dt_1.
\end{equation}
The last one is the stochastic part of the membrane potential:
$$
\Vknoise{t,\omega} = \frac{\sigma_B}{C_k} \int_{\tko}^{t}  \Gamma_k(t_1,t,\omega) \, dB_k(t_1),
$$
where $B_k(t_1)$ is a Brownian process, thus Gaussian.

\subsection{Gibbs distribution}

We are interested in the statistic of spikes generated by this model, in spontaneous activity and in the presence of the stimulus. We want to propose a linear response theory in this context. For this we first show that spike statistics is associated to a form of Gibbs distribution. 

\subsubsection{Transition probabilities}

Here, we want to compute the probability to have a spiking pattern $\omega(n)$ given the history, something informaly reading like $\Pnc{\omega(n)}{H_{<n}}$, where $H_{<n}$ is the history anterior to $n$, depending on voltage and spikes history. To simplify this dependence, we are going to make the approximation that the spike pattern $\omega(n)$ depends only on the spike history. This makes sense if one wants to use this theoretical approach to analyze experimental data on spike trains recordings with Multi-Electrode Arrays for example \cite{}. Here, indeed, one has only access to spikes history, not to voltage\footnote{MEA records actually a voltage activity on recording electrodes. Then, a spike sorting algorithm is used to attribute recorded voltage traces to neurons, and then, construct the spike train. Therefore, MEA could in principle allow to obtain the probability of having a spike pattern given the voltage history.}.

\begin{figure}
\center
\includegraphics[width=8cm,height=5cm]{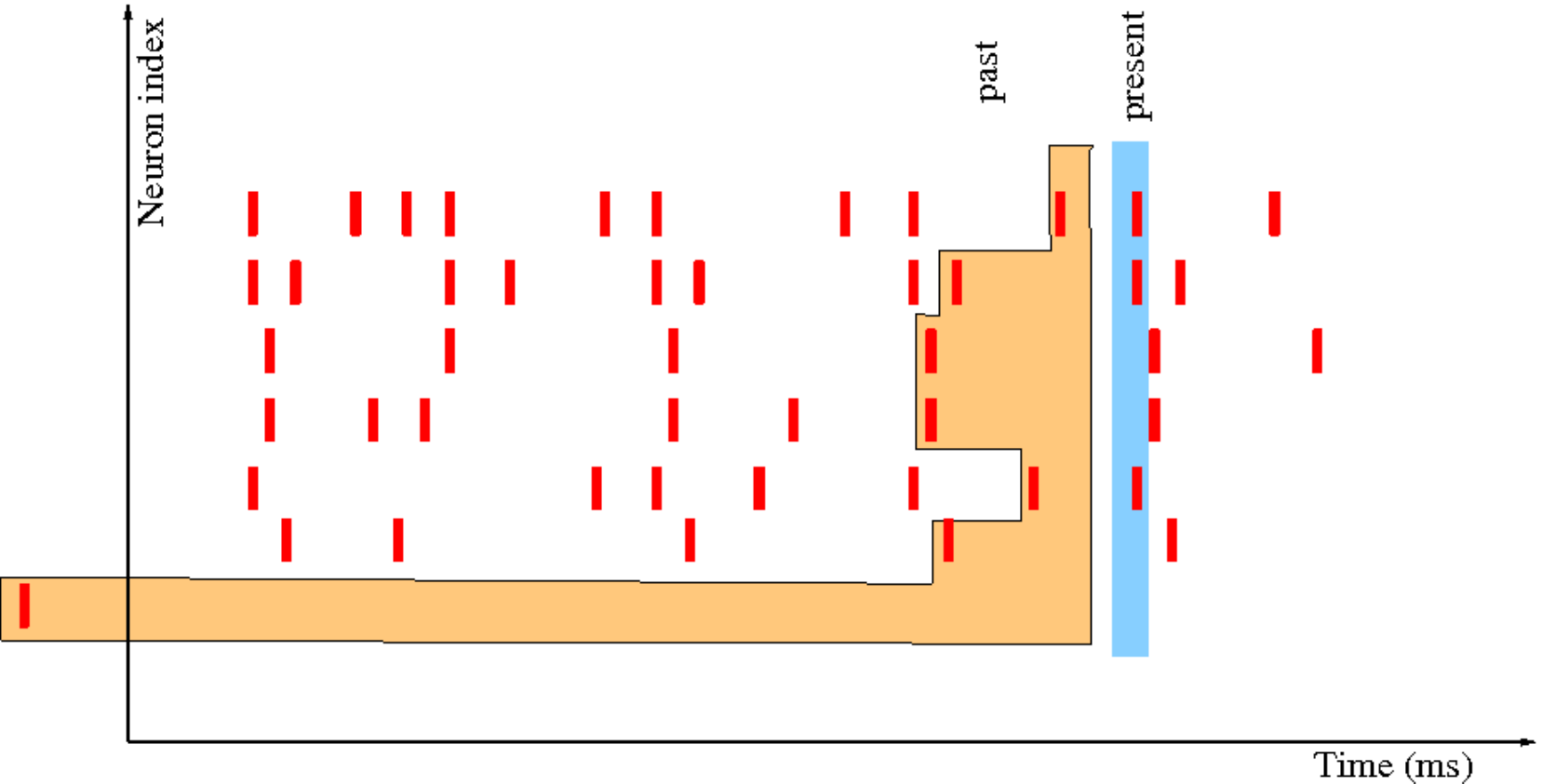}
\caption{Conditional probability of a spiking pattern (in blue) given the history. In Integrate and Fire models voltage memory (in orange) is reset when the neuron spikes (in red), so memory goes back up to the last time anterior to $n$ where the neuron has spiked (red spike in the orange area). This time is variable and can extend quite far in the past, giving rise to variable length Markov chain.\label{fig:memory}}
\end{figure}

Under this assumption $H_{<n}$ identifies with $\sif{n-1}$, where the memory extends in principle to the far past (here $-\infty$). This is an important point. In Integrate and Fire models  voltage memory is reset when the neuron spikes, so memory goes back up to the last time anterior to $n$ where the neuron has spiked. However, this time can be quite far in the past, giving rise to variable length Markov chain, as illustrated in Fig. \ref{fig:memory}. 
In addition, the spike memory does not only depend on voltage, it is also depends on conductances, as expressed in \eqref{eq:gk}, so memory is, in principle, infinite. We are therefore seeking probabilities of the form $\Probct{n}{\omega(n)}{\bloc{-\infty}{n-1}}$, where the subscript $n$  expresses that these probabilities depend on general on the discrete time $n$.

One can show\cite{cessac:11b} that these transition probabilities are well approximated by:
\beq\label{eq:Pjoint}
\Pnc{\omega(n)}{\sif{n-1}} = \prod_{k=1}^N
\Pnc{\omega_k(n)}{\sif{n-1}},
\eeq
where:
\begin{widetext}
\beq\label{eq:PMarg}
\Pnc{\omega_k(n)}{\sif{n-1}}=
\omega_{k}(n) \,  \Pi\pare{\frac{\theta-\Vkdet{n-1,\omega}}{\sk{n-1,\omega}}} \, + \, \pare{1-\omega_{k}(n)} \,
\pare{1-\Pi\pare{\frac{\theta-\Vkdet{n-1,\omega}}{\sk{n-1,\omega}}}},
\eeq
\end{widetext}
where $\Vkdet{n-1,\omega}$ is the deterministic part of the voltage,  given by \eqref{eq:Vk_integ} and:
$$
\skd{n-1,\omega}
= \left(\frac{\sigma_B}{C_k}\right)^2 \,
\int_{\tau_k(n-1,\omega)}^{n-1} \Gamma^2_k(t_1,n-1,\omega)  \, dt_1,
$$
corresponds to the  variance of the noise integrated along the flow up to time $n-1$.
Finally
$$
\Pi(x)=\frac{1}{\sqrt{2\pi}}\int_x^{+\infty} e^{-\frac{u^2}{2}}du.
$$
%
 

\subsubsection{Gibbs distribution}\label{eq:Gibbs}

Probabilities of this form define a generalisation of Markov chain called \textit{chains with complete connections}, a notion introduced by Onicescu and Mihoc in 1935 \cite{onicescu-mihoc:35}. The terminology chains with infinite memory can be also found. Under suitable  conditions \cite{fernandez-maillard:05,maillard:07} these transition probabilities define a probability $\mu$ on the set of spike trains which is a generalisation of the probability consistent with a Markov chain:
\beq\label{eq:def_consistent}
\begin{array}{lll}
&\int h\left(\seq{\omega}{-\infty}{n}\right) \mu(d\omega)\\ 
&= \int \sum_{\omega(n) \in \cA} h\left(\sif{n-1} \omega(n) \right)\Pnc{\omega(n)}{\sif{n-1}} \mu(d\omega).
\end{array}
\eeq
with $\cA=\Set{0,1}^N$. This equality must hold for all $n \in \mathds{Z}$ and measurable functions $h$. Obviously, the fact of conditioning by an infinite past requires some caution and conditions on the transition probabilities to ensure the existence of $\mu$.

 $\mu$ has  actually strong analogies with Gibbs distributions in rigorous statistical mechanics, where the set of spike trains can be viewed as a one dimensional spin chain labeled by the time index. An important difference, though, is that Gibbs distributions are constructed by conditioning upon left and right boundary conditions. In contrast, in the present context we only condition upon the past (left specification). For this reason, $\mu$ is rather called a Left Interval Specification (LIS) \cite{fernandez-maillard:05,leny:08}. There are mathematical examples showing that LIS have different properties than left-right conditionned Gibbs distribution \cite{fernandez-gallo:11}. However, to the best of my knowledge there is no effective, operational way, to distinguish these two notions from a finite sample obtained e.g. from numerical simulations, as used most of the time in the field of computational neuroscience. As a consequence, I will not distinguish these two notions and call $\mu$ a Gibbs distribution. 
 
 There are standard theorems ensuring the existence and uniqueness of a Gibbs distribution in this sense. In our case, these theorems apply because of the exponential decay of conductances \eqref{eq:gk} which induces an exponential decay of memory, the continuity of the family of transition probabilities and the summability of their variations (see \cite{cessac:11b,cofre-cessac:13} for details). 
 
A consequence of the definition \eqref{eq:def_consistent} is that the probability of a spike block $\bloc{m}{n}$, given the past reads:
\begin{equation}\label{eq:CondProbBlocChain}
\Pnc{\bloc{m}{n}}{\sif{m-1}}=e^{\Phi(m,n,\omega)} = 
e^{\sum_{r=m}^n \phi\pare{r,\omega}}
\end{equation}
where:
\beq\label{eq:pCondphi}
\Phi(m,n,\omega)=\sum_{r=m}^n \phi\pare{r,\omega},
\eeq
and
\beq\label{eq:phi_def}
\phi\pare{n,\omega} \equiv \log \Pnc{\omega(n)}{\sif{n-1}},
\eeq 
so that $\phi\pare{n,\omega}$ has the form of a Gibbs energy with infinite range. This "energy" being the log of a probability, there is no need to normalise with a partition function. We call $\phi$ a normalised energy (although it does not have the physical dimension of an energy, it is rather an information - bit rate). 
It depends explicitly on the model parameters via eq. \eqref{eq:Vk_integ}. It contains, as well, the stimulus influence, via the term \eqref{eq:Vkext}. We note  $\phisp{\omega}$ the energy in spontaneous activity, $\epsilon=0$ and $\musp$ the corresponding Gibbs distribution. It does not depend on $n$ as dynamics and transition probabilities are stationary in this case.  
$\phisp{\omega}$ plays therefore the role of energy in the equilibrium Gibbs distribution \eqref{eq:GibbsDef}.
 Yet, it does not have the form \eqref{eq:HStatPhys}. We come to this point in section \ref{sec:ApproxRepIF} (eq. \eqref{eq:HCdecom}).

\subsection{Linear response} \label{sec:LinRepIF}

\subsubsection{General form} \label{sec:GenFormRepIF}
Assume now that the spiking neuronal network receives  a time-dependent stimulus $S(t)$ from time $t_0$ to time $t_1$. Even if the stimulus is applied to a subset of neurons in the network, its influence will eventually propagate to other neurons, directly or indirectly connected. The stimulus will act on spikes timing, modifying spike correlations. 
 We note $n_0 = \ent{t_0}$, the integer part of $t_0$. For times anterior to $n_0$, $\mu$ identifies with $\musp$, that is, for any $m < n \leq n_0$, for any block $\bloc{m}{n}$, $\moy{\bloc{m}{n}}=\moysp{\bloc{m}{n}}$. In contrast, for $n>n_0$ spike statistics is modified. 

We consider a function (observable) $f(t,\omega)$. How is its average modified by the application of the stimulus ? We set $\Expm{\mu}{f(t,.)} \deq \Expm{\musp}{f(t,.)} + \delta \bra{f(t,.)}$ where $\delta \bra{f(t,.)} =0$ for $t<t_0$ and $\delta \bra{f(t,.)} \neq 0$ for $t \geq t_0$. We want to compute $\delta \bra{f(t,.)}$ when $\epsilon$, the stimulus amplitude, is weak enough.
Using a formal expansion of the energy \eqref{eq:phi_def} one obtains $\delta \mu\bra{f(t)}$ in the classical convolution form \cite{cessac-cofre:19}:
\begin{equation}\label{eq:RepConvIF}
\delta \mu\bra{f(t)}  \, = \epsilon \, \bra{\kappa_f * S}\pare{t},
\end{equation}
where the convolution kernel takes the form:
\begin{widetext}\begin{equation}\label{eq:kernelConv}
\kappa_{k, f} \pare{t-t_1}
\, = \,  \frac{1}{C_k} \, \sum_{r=-\infty}^{\ent{t-t_1}} \, C^{(sp)}\bra{f(t-t_1,.), 
 \frac{\mathcal{H}_k^{(1)}(r,.) }{\sk{r-1,.}}\,   \Gamma_k(0,r-1,.)}.
\end{equation}
\end{widetext}
Here the function $\mathcal{H}_k^{(1)}$ is obtained via a first order expansion of \eqref{eq:phi_def}. It contains therefore in particular the synaptic weights \eqref{eq:Wkj}.
The notation $C^{(sp)} \bra{A(t,.) B(s,.)}$ signifies the correlation of the functions $A(t,\omega)$ and $B(s,\omega)$ where the averaging is done with respect to the spike train distribution, in spontaneous dynamics, $\musp$· Thus, the linear response kernel is here as well obtained as a sum of correlation functions computed with respect to the spontaneous dynamics. Note that the kernel depends
on the flow of the dynamics $\Gamma_k(0,r-1,.)$  and on noise (term $\sk{}$). As in the equation \eqref{eq:Chi_Discrete_Chaos} computed in section \ref{sec:LinRepChaos}, the linear response kernel is obtained by a summation on the whole spike history. 

\subsubsection{Approximations} \label{sec:ApproxRepIF}

\paragraph{Markovian approximation.}

The exponential decay of the synaptic response $\alpha$ in \eqref{eq:alphakj} entails the existence of a time scale, $\tau$, after which the dependence in the past is essentially lost. This suggests to use a Markovian approximation where the transition probabilities $\Probc{\omega(n)}{\sif{n-1}}$ with infinite memory are replaced by $\Probc{\omega(n)}{\bloc{n-D}{n-1}} >0$, with a fixed memory depth $D>0$. Dynamics is then given by a Markov chain and $\musp$ is the invariant probability of this chain. The memory depth $D$ of the chain is constrained by the characteristic times:
\begin{equation}\label{eq:taudk}
\tau_{d,k} = \frac{C_k}{g_L+ \tau \sum_{j=1}^N G_{kj} \nu_j },
\end{equation}
where $\nu_j$ is the firing rate of neuron $j$ \cite{cessac-cofre:19}.\\

\paragraph{Hammersley-Clifford decomposition.} In the Markovian approximation the energy \eqref{eq:pCondphi} becomes a real function of spike blocks $\bloc{0}{D}$. A general theorem from Hammersley and Clifford\cite{hammersley-clifford:71}  states that the normalised energy $\phisp{\omega}$ can then decomposed on a basis of interaction functions:
\begin{equation}\label{eq:HCdecom}
\phisp{\omega} = \sum_{l} \,  \phi_l \, m_l\pare{\omega}
\end{equation}
where:
\begin{equation}\label{eq:monomial_def}
m_l(\omega)=\prod_{k=1}^n \omega_{i_k}(t_k).
\end{equation}
where $i_k = 1 \dots N$ is a neuron index, and $t_k=0 \dots D$. Thus,  $m_l(\omega)=1$ if and only if, in the spike train $\omega$, neuron $i_1$ spikes at time $t_1$, $\dots$, neuron $i_k$ spikes at time $t_k$. Otherwise, $m_l(\omega)=0$. The number $n$ is the degree of the interaction; degree one interactions have the form $\omega_{i_1}(t_1)$, degree $2$ interactions have the form $\omega_{i_1}(t_1)\omega_{i_2}(t_2)$, and so on. These interactions involve in general a time delay between spikes.

The form \eqref{eq:HCdecom} bares analogy with the energy form \eqref{eq:HStatPhys} where the interactions $m_l$ play the role of the $X_\alpha$s. More generally, an energy of the form:
\begin{equation} \label{eq:EnergyIF}
H\pare{\omega} = \sum_{\alpha} \lambda_\alpha \, X_\alpha\pare{\omega}
\end{equation}
where the $X_\alpha\pare{\omega}$s are functions of blocks $\bloc{0}{D}$ is associated to a Markov chain via the Perron-Frobenius theorem \cite{seneta:06,gantmacher:59}. Under moderate assumptions ($\lambda_\alpha$ are bounded from below), this chain has a Gibbs invariant probability \cite{cofre-cessac:14}.

In the simplest case, ($D=0$, pairwise interactions) this probability is the Gibbs distribution of an Ising model \cite{cofre-cessac:14}. This has attracted much interest in the computational neuroscience community although the Ising form is just the simplest non trivial potential existing in this context. We come back to this point below.\\

\paragraph{Further approximations.}
It is possible to simplify the form \eqref{eq:kernelConv} with the following approximations:
\begin{enumerate}[(i)]
\item Replace $\nkrm$  by $-\infty$; 
\item Replace $\Gamma_k(t_1,r-1,\omega)=e^{-\frac{1}{C_k}\int_{t_1}^{r-1}\gk{u,\omega} \, du}$ by  $e^{-\frac{(r-1-t_1)}{\tau_{d,k}}}$.
\end{enumerate}

Then, linear response reads:
\begin{widetext}
\begin{equation}\label{eq:LinRepExpansionIF}
\dmu{f(t)} \, =
- \frac{2}{\sigma_B}  \sum_{k=1}^N \frac{1}{\sqrt{\tau_{d,k}}}  \sum_{r=-\infty}^{n= \ent{t}}  
\bra{
\begin{array}{lll}
&&\\
&&\\
&&\\
\end{array}
}
 \pare{S_k \ast e_{d,k}}(r-1)
\end{equation}
\end{widetext}
where:
$$
e_{d,k}(u) = e^{-\frac{u}{\tau_{d,k}}}
$$
and where the term $\bra{
\begin{array}{lll}
&&\\
&&\\
&&\\
\end{array}
}$ is an expansion of correlation functions between the observable $f$ and spikes interactions. In the case of a memory depth $D=1$ it reads :
\begin{widetext}
\begin{equation} \label{eq:ExpansionBracketIF}
\bra{
\begin{array}{lll}
&&\\
&&\\
&&\\
\end{array}
}
=
\gamma_k^{(1)} \,  \Csp{f(t,\cdot)}{\omega_k(r)}\\
+\sum_{i=1}^N \gamma_{k;i}^{(2)} \, \Csp{f(t,\cdot)}{\omega_k(r) \,\omega_i(r-1) 
}
+\sum_{i,j=1}^N \gamma_{k;ij}^{(3)} \, \Csp{f(t,\cdot)}{\omega_k(r) \,\omega_i(r-1) \,\omega_j(r-1)
}
+ \cdots
\end{equation}
\end{widetext}
where correlations are computed with respect to the equilibrium probability. For example, the variation induced by the stimulus, in the firing rate of neuron $m$, at time $t$ is given by:
\begin{widetext}
\begin{equation} \label{eq:ExpansionrateIF}
\begin{array}{lll}
\dmu{\omega_m(t)} \, =\\
- \frac{2}{\sigma_B}  \,\sum_{k=1}^N \, \frac{1}{\sqrt{\tau_{d,k}}} \, \sum_{r=-\infty}^{n= \ent{t}}  
\bra{
\begin{array}{ll}
&\gamma_k^{(1)} \, \Csp{\omega_m(t)}{\omega_k(r)}\\
+\sum_{i=1}^N &\gamma_{k;i}^{(2)} \, \Csp{\omega_m(t)}{\omega_k(r) \,\omega_i(r-1) 
}\\
+\sum_{i,j=1}^N &\gamma_{k;ij}^{(3)} \, \Csp{\omega_m(t)}{\omega_k(r) \,\omega_i(r-1) \,\omega_j(r-1)
}\\
+& \cdots
\end{array}
}
\, \pare{S_k \ast e_{d,k}}(r-1)
\end{array}
\end{equation}
\end{widetext}

More generally, it involves correlations to pairwise, triplets, etc spike time correlations.
The coefficients $\gamma_k^{(l)}$ depends on the synaptic weights $W_{kj}$ and are therefore constrained by neurons interactions.

\subsection{Conclusions of section \ref{sec:LinRepSpiking}}\label{sec:ConclusionsIF}

In this section, we have derived a linear response for a spiking neural network. As for the Amari Wilson Cowan model we end up with a convolution kernel depending on synaptic graph and equilibrium correlations. Yet, the form obtained is quite more complex than the form \eqref{eq:Chi_Discrete_Chaos}, and quite harder to interpret. The interesting point is that the terms of this expansion, which are spike-time correlations computed at equilibrium, can be obtained by an ergodic, time average, in spontaneous activity. The numerical computation of high order terms is, however, cumbersome and will be the subject of a forthcoming paper. 

As in section \ref{sec:LinRepFiring} the correlations decay exponentially fast, so that it is possible to truncate the expansion \eqref{eq:LinRepExpansionIF}
to low orders, e.g. pairwise interactions for neurons, and small memory depth, as e.g. eq. \eqref{eq:ExpansionrateIF}. Yet, many terms remain. 
At the lowest non trivial order the spontaneous energy contains only synchronous pairwise interactions of the form $\omega_i(0)\omega_j(0)$ and the energy \eqref{eq:EnergyIF} corresponds to a Ising model, where successive times are independent. An expansion similar to \eqref{eq:LinRepExpansionIF} has be done by S. Cocco et al \cite{cocco-leibler-etal:09}. The Ising model has been used by many authors to analyze spike trains statistics, especially in retina data \cite{schneidman-berry-etal:06,shlens-field-etal:06,tkacik-schneidman-etal:09,tkacik-mora-etal:15,gardella-marre-etal:19}. It neglects, however, higher order correlations which have been shown to play a role in spike response 
to stimuli \cite{ganmor-segev-etal:11,ganmor-segev-etal:11b,vasquez-marre-etal:12}, e.g. in the retina. \\

This work is at a less advanced stage than the previous one and there remain some pending questions:

\paragraph{Resonances.} Submitting this spiking network to an harmonic stimulus, one may observe resonances, like in the Amari-Wilson-Cowan model. In the Markovian case (finite memory with depth $D$) these resonances are the eigenvalues of the transition matrix defined through the transition probabilities $\Probc{\omega(n)}{\bloc{n-D}{n-1}}$. They are difficult to compute directly though, even numerically, because the dimension of the matrix growths exponentially fast with the number of neurons and the memory depth $D$. Analytic computations seem out of reach. Therefore, resonances would have to be computed the same way as in section \ref{sec:NumericsLinRepChaos}, eq. \eqref{eq:Chi_ij_Numerical}. We expect the same observations as for the Amari-Wilson-Cowan model: frequency dependent effective connectivity and frequency dependent clusters of  synchronising neurons. 

\paragraph{Structure of spike correlations.} One of the goal guiding the work presented in this section is to understand how the spatio-temporal correlations present in the trajectory of a moving object are reflected in a spiking neuronal network when this moving object is the stimulus. Cells are excited by the moving object, but, as they are connected, this excitation triggers a wave of activity in the network which can for example enhance the effect, inducing anticipation \cite{souihel-cessac:19}.  We are essentially interested in applying this formalism to understand how the retina encodes motion, and how can one decode this motion for the spatio-temporal correlations observed in the output retinal cells (Ganglion cells).

\section{Conclusion}\label{sec:Conclusion}

In this paper we have considered, at a modelling level, the effects induced on a neuronal network by the  weak stimulation of a sub group of neurons, using linear response theory. The goal was actually twofold: (i) analyse this effect starting from the equations ruling the neurons dynamics; (ii) make a link to linear response in non equilibrium statistical physics . For this we borrowed existing results, either coming from ergodic theory and chaotic dynamical systems, or from Markov chain and chains with complete connections. In the two proposed examples we ended up in a relation expressing the linear response to the stimulus as a convolution of this stimulus with a kernel. The form of the kernel is complex: as for statistical physics (e.g. kinetic theory) it depends on the microscopic dynamics, on the interactions (in our case, the synapses) and it is obtained via a suitable averaging that smooths the microscopic trajectory and establish a description at a mesoscopic level. The averaging, performed with respect to the equilibrium distribution, corresponds to an ergodic time-average, with no necessity to consider a thermodynamic limit where the number of neurons tends to infinity. 

The analogy with statistical physics can go relatively far allowing us to define susceptibility, currents and Green-Kubo like relations. However, one of the main difficulty here is to obtain \textit{a priori} the correct form for the "energy" \eqref{eq:HStatPhys}. In physics this form is guided by first principles, mechanics or thermodynamics. What could be the equivalent principles in neuroscience, if any ? At the moment we are faced to two possible strategies: Either use formal analogies with statistical physics, e.g. proposing Ising model as a canonical or lowest order normal form for the energy, or try and extract the energy form from microscopic dynamics. The first approach raises the question of the role played by higher order terms and the way how to characterise their effects. The second approach resembles the program initiated by Boltzman to fund thermodynamics from mechanics, a program far from being completed yet, despite deep progresses \cite{gallavotti:96}. Finally, the pending question is: why should statistical physics give fruitful insight in the understanding of neuronal dynamics and, more generally, neuroscience ? Beyond the fact that this conceptual analogy has lead to interesting new concepts in neuroscience, like in Friston's theory \cite{friston-kilner-etal:06}, the answer lies maybe above statistical physics, in large deviations and Markov chains \cite{kaiser-jack-etal:18}. 

This is actually the hidden link between section \ref{sec:LinRepFiring} and section \ref{sec:LinRepSpiking}. Indeed, a way to define the SRB state and to derive a linear response theory for chaotic systems is to use a salient property of chaotic (uniformely hyperbolic-like) dynamical systems. They have Markov partitions and can be encoded by Markov chains. Hyperbolicity allows indeed to split the phase space into a finite partition, constructed from local pieces of stable and unstable manifolds, and to encode the dynamics by a Markov transition matrix between the elements of this partition. The transition probabilities of this chain  are weighted by the exponential of the potential \eqref{eq:PotentielSRB}. The SRB state is the invariant probability  of this chain, while the induced large deviations of the entropy production allows to define currents and Onsager coefficients \cite{gallavotti-cohen:95,gallavotti-cohen:95b,gallavotti:96}. 

An interesting alternative approach, that could be fruitful when applied to neuronal models, has been developed by C. Maes in co-workers in \cite{baiesi-maes:13,basu-maes:15}. It is based on dynamical operator (also called Koopman operator \cite{gaspard:05}) acting on observables, whose adjoin is the Ruelle-Perron-Frobenius operator. This approach somewhat synthesises and generalises previous approaches giving a fast and compact way to obtain the linear response and higher order terms. It would be interesting to apply this method to neuronal network although it might be difficult to obtain a tractable form of the evolution operator.

More generally, the main obstacle to apply methods coming from non equilibrium statistical physics
is that they have mainly be developed for physical problems and most examples are guided by physics and thermodynamics. In contrast, we don't have such guidelines here. We have no a priori idea of what is the form of the energy, and thermodynamics principles are useless to understand dynamics; for example there is no notion of "heat" or "work" that would allow to understand \eqref{eq:AWC} or \eqref{eq:subthresholdVk}. The "thermodynamics of the brain" is still under investigation \cite{kirkaldy:65}.\\

Returning to linear response, we would like to address several questions left aside in the paper. \\

\paragraph{Beyond linear response.} In this paper, the linear response has been derived from an expansion in $\epsilon$, the amplitude of the perturbation. What about the higher order terms that we have neglected ?
In the context of chaotic systems higher order terms have been computed by Ruelle\cite{ruelle:98} and are therefore accessible mathematically in the model \eqref{eq:AWC_Discrete}. Their form is quite hard to interpret in our context though, in contrast to the simple first order term of eq. \eqref{eq:Chi_Discrete_Chaos}. They are also very hard to estimate numerically or experimentally. 
It is known from Volterra expansion theory that higher order terms are given by high order correlations, which are quite difficult to estimate experimentally, requiring very large samples. We are confronted here to a similar problem. 

If we stick at the questions addressed in this paper, the $\epsilon$-expansion approach does not seem reasonable beyond the first order. In the field of neuroscience researchers prefer to correct the linear response convolution by a static non linearity \cite{rieke-warland-etal:97}. In my opinion, the main interest of linear response theory in the context of neuronal modelling is to give us a notion of derivative of a statistical quantity (the average of an observable) with respect to a time dependent perturbation, in terms of the dynamics ruling the evolution of neurons. This leads, in the case studied here, to an explicit form for the convolution kernel, especially how it depends on dynamics and synaptic connections. In this setting, one sees explicitly that the "response" of a cell is not intrinsic to that cell, but depends on its dynamical surrounding. This approach could be useful to infer the receptive field of sensory cells from the structure of the neuronal circuits they are involved in. For example, it is known, in the retina, that the center-surround receptive field structure of bipolar and ganglion cells is due
to the lateral inhibitory connectivity of horizontal cells. More complex circuits, processing motion, exist as well in the retina, involving amacrine cells whose role is far from being understood yet \cite{gollisch-meister:10}. The present approach could afford to anticipate the role played by these cells in shaping the receptive field of cells responding to motion features.  \\

\paragraph{Effective interactions.}
Linear response lead us to define a notion of effective interactions from the underlying non linear dynamics. As we argued, these interactions are not the synaptic connections, and they do not reduce to correlations. It is commonly accepted in the neuroscience community that "information" is transported by neurons. This is characterised by mutual information, relative entropy,  Granger causality, ... In this paper we came out with the proposal \eqref{eq:Currentj-i}, still on a shaky ground, based on the non linear effects induced by the sigmoid and a notion of entropy production. A next step is to investigate how to compute this quantity and how it compares with standard indicators.\\

\paragraph{Closeness to bifurcations.} Neurons can exhibit drastic changes (bifurcations) in their behaviour when they are stimulated with a stimulus of increasing amplitude. The same holds as well in a neuronal network. If some neurons are close to a bifurcation point, a tiny stimulation of a single neuron can drastically change its activity, which, in turn, can induce bifurcations of other neurons in an avalanche like or wave activity. Thus, a small perturbation leads to a macroscopic change (diverging susceptibility). Such mechanism play an important role in neuronal dynamics. It is a current belief, observed in models, that learning and plasticity drive neuronal networks close to bifurcations points where they respond fast and optimally to learned stimuli\cite{bertschinger-natschlager:04,siri-berry-etal:07,siri-berry-etal:08,naude-cessac-etal:13}.
On mathematical grounds,  near bifurcations point, there is a loss of structural stability that can ruin any hope to have a linear response theory although structural stability is not necessary to obtain linear response and can be extended near bifurcations point under some conditions \cite{baladi-benedicks-etal:14}.  

Linear response can be useful though to characterise the approach to such critical points. For example, the divergence of susceptibility could correspond to resonances converging to the real axis, in a Lee-Yang like phenomenon \cite{yang-lee:52}, providing a strong analogy between the behaviour of neuronal networks near bifurcations and phase transitions.

\begin{acknowledgements}
I would like to acknowledge Jacques-Alexandre Sepulchre and Rodrigo Cofr\'e from the good time we spend together thinking and working on the linear response results presented in this paper. 
I am grateful to Maria-José Escobar, Patricio Orio, Rodrigo Cofr\'e and Wael El-Deredy to give me the opportunity to teach this lecture in the LACONEU summer school 2019. I would like to acknowledge the students of this school for their questions and their enthusiasm.

I thank the Reviewers for their insightful comments and constructive criticism.
\end{acknowledgements}




\end{document}